\documentclass[aps,prl,reprint,superscriptaddress,floatfix]{revtex4-2}
\usepackage{amsmath,amssymb,wasysym}
\usepackage{graphicx}
\usepackage{dcolumn}
\usepackage{bm}
\usepackage{braket}
\usepackage{verbatim}
\usepackage{float}
\usepackage{bbold}
\usepackage{color}
\usepackage{xcolor}
\usepackage{relsize}
\usepackage{amsthm}
\usepackage{enumerate}
\usepackage{soul,xcolor}
\usepackage[T1]{fontenc}
\usepackage{adjustbox}
\usepackage[colorlinks=true ,urlcolor=blue,urlbordercolor={0 1 1}]{hyperref}

\newcommand{\be}{\begin{equation}}
\newcommand{\ba}{\begin{align}}
\newcommand{\ee}{\end{equation}}
\newcommand{\bea}{\begin{eqnarray}}
\newcommand{\eea}{\end{eqnarray}}
\newcommand{\beq}{\begin{equation}}
\newcommand{\eeq}{\end{equation}}
\newcommand{\beqn}{\begin{eqnarray}}
\newcommand{\eeqn}{\end{eqnarray}}

\renewcommand{\hat}[1]{{\widehat #1}}

\usepackage{bm}
\usepackage{array}
\newcolumntype{L}[1]{>{\raggedright\arraybackslash}p{#1}}
\newcolumntype{C}[1]{>{\centering\arraybackslash}p{#1}}
\newcolumntype{R}[1]{>{\raggedleft\arraybackslash}p{#1}}
\usepackage{multirow}
\allowdisplaybreaks

\begin{document}
\title{High spin, low spin or gapped spins: magnetism in the bilayer nickelates}

\author{Hanbit Oh}
\thanks{These two authors contributed equally}
\affiliation{William H. Miller III Department of Physics and Astronomy, Johns Hopkins University, Baltimore, Maryland, 21218, USA}
\author{Yi-Ming Wu}
\thanks{These two authors contributed equally}
\affiliation{Stanford Institute for Theoretical Physics, Stanford University, Stanford, California 94305, USA}
\author{Julian May-Mann}
\affiliation{Department of Physics, Stanford University, Stanford, CA 94305, USA}
\author{Yijun Yu}
\affiliation{Stanford Institute for Materials and Energy Sciences, SLAC National Accelerator Laboratory,
Menlo Park, CA 94025, USA}
\affiliation{Departments of Applied Physics, Stanford University, Stanford, CA 94305, USA}
\author{Harold Y. Hwang}
\affiliation{Stanford Institute for Materials and Energy Sciences, SLAC National Accelerator Laboratory,
Menlo Park, CA 94025, USA}
\affiliation{Departments of Applied Physics, Stanford University, Stanford, CA 94305, USA}
\author{Ya-Hui Zhang}
\email{yzhan566@jhu.edu}
\affiliation{William H. Miller III Department of Physics and Astronomy, Johns Hopkins University, Baltimore, Maryland, 21218, USA}
\author{S. Raghu}
\email{sraghu@stanford.edu}
\affiliation{Stanford Institute for Theoretical Physics, Stanford University, Stanford, California 94305, USA}
\affiliation{Stanford Institute for Materials and Energy Sciences, SLAC National Accelerator Laboratory,
Menlo Park, CA 94025, USA}

\date{\today}

\begin{abstract}
Inspired by the recent discovery of high-temperature superconductivity in bilayer nickelates, we investigate the role of magnetism emerging from a hypothetical insulating $d^8$  parent state. We demonstrate that due to the interplay of superexchange and Hund's coupling, the system can be in a high-spin, low-spin or spin-gapped state. The low-spin state has singlets across the bilayer in the $d_{z^2}$ orbital, with charge carriers in the $d_{x^2-y^2}$ orbital.  Thus, at low energy scales, it behaves as an effective one band system when hole doped. By contrast, the high-spin state is a more robust, spin-1 antiferromagnet. Using Hartree-Fock methods, we find that for fixed interaction strength and doping,  high-spin magnetism remains more robust than the low-spin counterpart. Whether this implies that the high spin state provides a stronger pairing glue, or more strongly competes with superconductivity remains an open question. Our analysis therefore underscores the importance of identifying the spin state for understanding superconductivity in nickelates.
\end{abstract}

\maketitle

{\it Introduction ---} Soon after the discovery of high temperature superconductivity (HTS) in a family of bilayer nickelates -- first in pressurized bulk crystals~\cite{sun2023signatures,HouJun_2023,PhysRevX.14.011040,Yuanhuiqiu2024}, and later in compressively strained films~\cite{Ko2025,Zhou2025,LiuYidi2025,hao2025superconductivityphasediagramsrdoped} --  intense effort has been devoted to understanding the emergent properties of these systems.  Experiments have begun to shed some light on structural properties~\cite{Wang2024,bhatt2025resolvingstructuraloriginssuperconductivity}, electronic band structure~\cite{Yang2024,Li_2025,shen2025anomalousenergygapsuperconducting,wang2025electronicstructurecompressivelystrained} and magnetic orders proximate to superconductivity~\cite{Chen2024,XIE20243221,PhysRevLett.132.256503,plokhikh2025unravelingspindensitywave,Zhao_2025,gupta2024anisotropicspinstripedomains,RenXiaolin2025}.  On the theoretical front, an impressive collection of works have already provided detailed proposals for the electronic structure, correlations and their implications for superconductivity~\cite{PhysRevLett.131.126001,PhysRevLett.132.146002,PhysRevB.108.174511,PhysRevB.110.104517,Liu2024,Yang2024,ZHANG2024147,PhysRevB.108.L140505,PhysRevB.111.174506,shen2023effective,Wú2024,PhysRevB.111.L020504,PhysRevLett.131.206501,PhysRevLett.131.236002,PhysRevB.109.L081105,lu2023superconductivity,PhysRevLett.132.126503,PhysRevB.109.165154,PhysRevLett.133.126501,PhysRevB.108.L140504,PhysRevB.108.174501,PhysRevB.108.165141,PhysRevB.108.L201108,2023arXiv230812750K,fabian1,fabian2,fabian3,zhan2024cooperation,PhysRevMaterials.8.074802,wang2024pressure,Xue_2024,Zhang2024,jiang2024high,PhysRevB.108.214522,oh2025dopingspinonemottinsulator,wang2025originspinstripesbilayer}.  

In the wake of a new experimental discovery, there is the temptation towards classification, {\it i.e.} to ascertain whether the new phenomena belongs to a category of its own, or represents an interesting variation on a known theme.  While HTS in the thin film bilayer nickelate is reminiscent  of the overdoped cuprates~\cite{liu2025superconductivitynormalstatetransportcompressively},  there remain obvious distinctions, the most prominent one being the discrepancy in valence count. For example, in the La-based cuprate parent compound La$_2$CuO$_4$, the low energy behavior is determined by the copper $d^9$ configuration, in which electronic excitations stem from the $d_{x^2-y^2}$ orbital and oxygen $p$ orbitals~\cite{Emery1987, Varma1987, PhysRevB.37.3759}.  By contrast, the $d^{7.5}$ valence in the bilayer nickelate La$_3$Ni$_2$O$_7$ necessitates the importance of multiple $d$-orbitals and Hund's coupling. To determine whether such differences in local quantum chemistry lead to new emergent behavior, it is useful to take a step back and ask questions that are more qualitative in nature.

In this letter, we demonstrate how the similarity (or lack thereof) between bilayer nickelates and cuprates crucially hinges on  the spin state in the nickelate bilayer. Based on simple quantum chemistry considerations,  we analyze magnetism in a putative parent insulating compound with $d^8$ electronic configuration.  Such a configuration may occur in compounds such as La$_2$CeNi$_2$O$_7$ or La$_3$Ni$_2$O$_{6.5}$.  We show that due to the interplay of Hund's coupling and superexchange, spins can take several configurations including spin-1 (``high spin''), spin-1/2 (``low spin''), or a gapped spin-0 configuration.  We argue that the low spin  configuration  is well-described by a 
single orbital bilayer Hubbard model.  Furthermore, the bilayer structure plays a crucial role in stabilizing this low spin state.  By contrast, the spin-1 and spin-0 configurations exhibit magnetic properties that are rather distinct from the cuprates. In the spin-1 configuration, the doped system is well-described by a ferromagnetic Kondo lattice. And recently, two of us have shown that the doped spin-0 state exhibits superconductivity that is quite different than the cuprates~\cite{yang2024strong,oh2024hightemperaturesuperconductivitykineticenergy}.  
Thus, it is of great importance to determine the spin moment in this material, which can be extracted from magnetic susceptibility, neutron scattering or X-ray scattering measurements.

\textit{Bilayer two-orbital Hubbard model ---} 
The bilayer nickelate consists of a $d^{7.5}$ configuration of Nickel electrons in a tetrahedral crystalline environment.  Consequently, the low energy electronic excitations come from partially filled, non-degenerate Nickel $d_{3z^2-r^2}$ (denoted as $``Z"$) and $d_{x^2-y^2}$ (denoted as $``X"$) orbitals. 
Thus, we begin with a two-orbital Hubbard model on a bilayer square lattice (see Fig.~\ref{fig:1}(a)), incorporating the $X,Z$ orbitals of Ni ions.  Owing to the shape and orientation of these orbitals, it follows that electrons in the $Z$-orbital tunnel predominantly {\it between} the Nickel planes of the bilayer, whereas electrons in the $X$-orbital primarily tunnel {\it within} these planes (see Fig~\ref{fig:1}(a)).  The Hamiltonian that incorporates only these  strongest orbital hybridization effects is
\begin{align}
H&= 
- t_\parallel \sum_{\langle i,j \rangle, \ell, \sigma} \left( d^\dagger_{i\ell\sigma} d_{j\ell\sigma} + \text{h.c.} \right) 
- t_\perp \sum_{i,\sigma} \left( f^\dagger_{i1\sigma} f_{i2\sigma} + \text{h.c.} \right) \nonumber \\
&\quad + U \sum_{i\ell} \left( n^d_{i\ell\uparrow} n^d_{i\ell\downarrow} + n^f_{i\ell\uparrow} n^f_{i\ell\downarrow} \right)
- J_H \sum_{i\ell} \vec{S}^d_{i\ell} \cdot \vec{S}^f_{i\ell},\nonumber \\
& \quad +\Delta \sum_{i,\sigma} [n_{i\ell\sigma}^d
-n_{i\ell\sigma}^f],
\label{eq:two_hubbard}
\end{align}
where $i$ denotes the site index and $\langle i,j \rangle$ run over nearest neighbor
pairs on the square lattice.
$\ell = 1,2$ labels the top and bottom layers, and $\sigma = \uparrow, \downarrow$ is the spin index. The operators $d^\dagger_{i\ell\sigma}$ and $f^\dagger_{i\ell\sigma}$ create electrons in the $X$ and $Z$  orbitals, respectively, with quantum numbers $(i,\ell,\sigma)$. 
$\Delta$ is an energy splitting between the two orbitals and tunes their relative density.
The number operators are defined as $n^d_{i\ell\sigma} = d^\dagger_{i\ell\sigma} d_{i\ell\sigma}$ and $n^f_{i\ell\sigma} = f^\dagger_{i\ell\sigma} f_{i\ell\sigma}$, while the spin 1/2 operators are $\vec{S}^d_{i\ell} = \frac{1}{2} d^\dagger_{i\ell\sigma} \vec{\sigma}_{\sigma\sigma'} d_{i\ell\sigma'}$ and similarly for $\vec{S}^f_{i\ell}$.
The Hamiltonian above does not include subdominant single electron tunneling processes such as the  interlayer hopping of $X$ electrons and in-plane hopping of $Z$ electrons. We also include the onsite Hubbard interaction $U$, taken for simplicity to be the same for each orbital,  and a ferromagnetic Hund’s coupling $J_H$ between the two orbitals on the same site. Given our interest here in magnetism, we take a strong coupling limit  $U/t_\parallel, J_H/t_\parallel \gg 1$.  

We define the hole concentration $x$ relative to a putative $d^8$ insulating configuration as follows: the total density in the $X$,$Z$-orbitals is  $n_T=2-x$.  The case of $x = 0.5$ corresponds to the experimentally relevant filling in La$_3$Ni$_2$O$_7$. The doping evolution is schematically illustrated in Fig.~\ref{fig:1}(b). We first analyze magnetism in the $d^8$ insulating system, and then proceed to consider the effects of hole doping $x>0$.

\begin{figure}[tb]
    \centering
\includegraphics[width=1.00\linewidth]{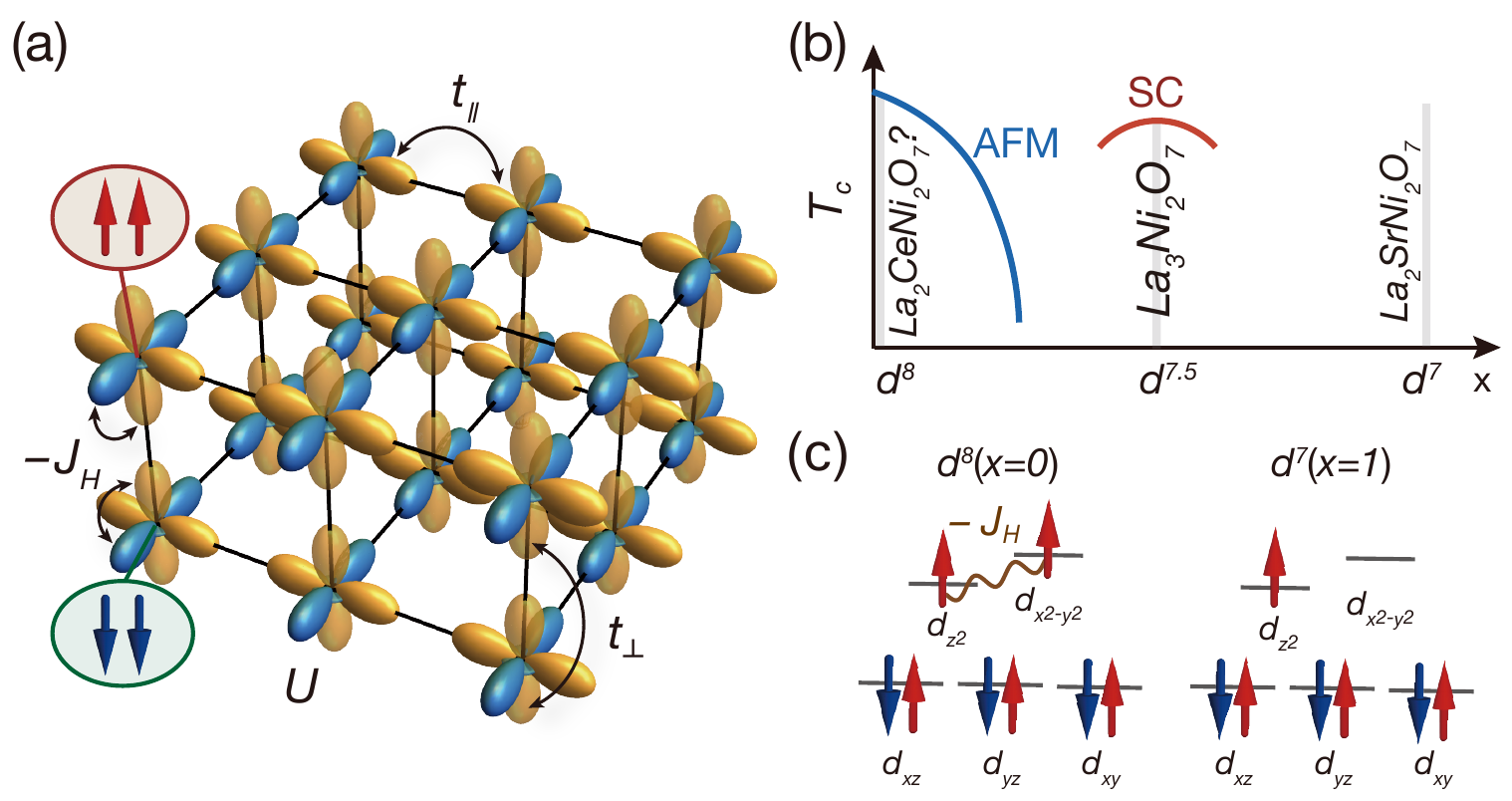}
    \caption{(a) Illustration of the bilayer two-orbital Hubbard model. The system consists of a bilayer square lattice with in-plane hopping $t_\parallel$ for $d_{x^2 - y^2}$ orbital and interlayer hopping $t_\perp$ for $d_{z^2}$ orbital. A ferromagnetic Hund’s coupling $J_H$ and an onsite Hubbard $U$ are included.
(b) Schematic phase diagram across hole doping from the hypothetical $3d^8$ parent state. La$_3$Ni$_2$O$_7$, with an average Ni valence of $d^{7.5}$, can be viewed as a 50\% hole-doped state from a $d^8$ Mott insulator. 
(c) Electronic configuration of Ni at $d^8$ and $d^{7}$.
}
    \label{fig:1}
\end{figure}

\textit{Magnetism in the insulating $d^8$ limit ---}  
At $x=0$, both the $X$ and $Z$-orbitals are half-filled and the model in Eq.~\eqref{eq:two_hubbard} reduces to an effective two orbital, spin-1/2 bilayer Heisenberg model (see Fig.~\ref{fig:2}):
\begin{align}
H_J &= 
J_\parallel \sum_{\langle i,j \rangle, \ell} \vec{S}^d_{i\ell} \cdot \vec{S}^d_{j\ell}
+ J_\perp \sum_i \vec{S}^f_{i1} \cdot \vec{S}^f_{i2}
- J_H \sum_{i\ell} \vec{S}^f_{i\ell} \cdot \vec{S}^d_{i\ell}
\label{eq:two_heisenberg}
\end{align}
There are three energy scales, $(J_\parallel,J_\perp,J_H)$. $J_\parallel = 4t_\parallel^2 / U$ and $J_\perp = 4t_\perp^2 / U$ are the in-plane and interlayer antiferromagnetic exchange couplings for the $X$ and $Z$-orbitals, respectively. We shall let $J_\parallel$ define a characteristic energy scale for the spin system and set $J_\parallel=1$ below.

\begin{figure}[b]
    \centering
\includegraphics[width=1.00\linewidth]{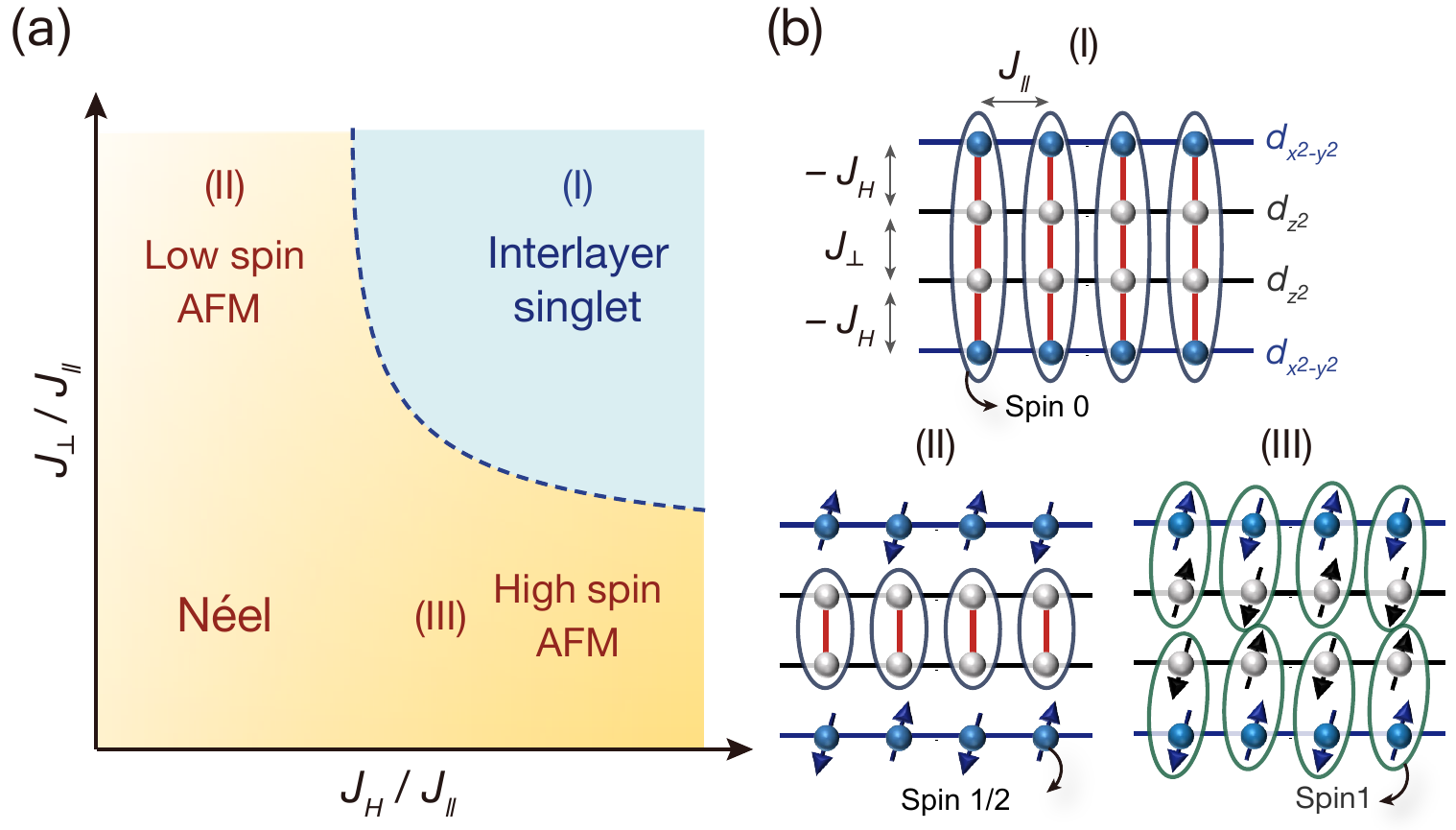}
    \caption{
(a,b) Schematic phase diagram and magnetic phases of $J_\parallel-J_\perp-J_H$ model, Eq.~\eqref{eq:two_heisenberg}.
(a) The phase diagram exhibits a second-order phase transition between a N\'eel-ordered phase at small $J_\perp$ and a interlayer singlet phase at large $J_\perp$. Within the N\'eel phase, a crossover occurs from a low-spin to a high-spin AFM as $J_H$ increases.
(b) Illustrations of the three phases: (I) interlayer singlet, (II) low-spin AFM, and (III) high-spin AFM. Each panel shows four horizontal layers representing two orbitals on each bilayer plane. 
Black ellipses with red lines in (I) and (II) indicate interlayer singlet bonds.
    }
    \label{fig:2}
\end{figure}

It is easy to deduce the ground state of this spin model in various asymptotic limits. 
When $J_H  \rightarrow 0$, the two orbital sectors are decoupled, and with $J_{\perp} \rightarrow  \infty$,  the ground state is a direct product of two independent spin systems: a N\'eel-ordered phase from spins in the $X$-orbitals and an interlayer singlet phase formed by $Z$-orbital spins directly across the bilayer. The $Z$ spins possess a spin-gap without any broken symmetry.  We shall dub this the ``low spin'' phase, as magnetism in this limit is dictated by spin-1/2 degrees of freedom.  In the opposite limit when $J_{\perp} \rightarrow 0$ and $J_H  \rightarrow \infty$, the spins in the two orbitals lock into a local spin-triplet configuration, forming an effective spin-1 moment per site. The system then maps onto a bilayer spin-1 Heisenberg model governed by $(J_\parallel, J_\perp)$.  We refer to this as the ``high spin'' phase.  Lastly, when both $J_H, J_{\perp} \rightarrow \infty$ with $J_{\perp} \gtrsim J_{H}$, the effective spin-1 degrees of freedom preferably form interlayer singlets, which results in a spin-gap without broken symmetry.  In this limit, there is no magnetism.  Thus, we expect a quantum phase transition between the spin-gapped interlayer singlet state at $J_H, J_{\perp} \gg 1$ to a N\'eel state when either $J_H$ or $J_{\perp} \sim \mathcal O(1)$.  Note that since both the high and low spin N\'eel states exhibit the same pattern of broken symmetry, there will be a crossover between these extremes, or at best a first order transition analogous to the liquid-gas transition.   See Fig.~\ref{fig:2} for a schematic phase diagram.

To arrive at more quantitative conclusions, we employ bond-operator mean field theory (BOMFT), which captures quantum disordered phases such as the interlayer singlet phase studied here.  Recent work~\cite{PhysRevB.84.214412} has shown that in the $J_H \rightarrow \infty$ limit of Eq.~\eqref{eq:two_heisenberg}, BOMFT quantitatively agrees with results from  quantum Monte Carlo (QMC) simulation. For completeness, we have also studied the phase diagram in the more standard Schwinger-Boson mean-field theory (SBMFT)~\cite{auerbach2012interacting,PhysRevB.38.316,PhysRevLett.66.1773}, which is in qualitative agreement with the BOMFT results and is relegated to the SM (See Fig.~S1).

\begin{figure}[t]
    \centering
\includegraphics[width=1.0\linewidth]{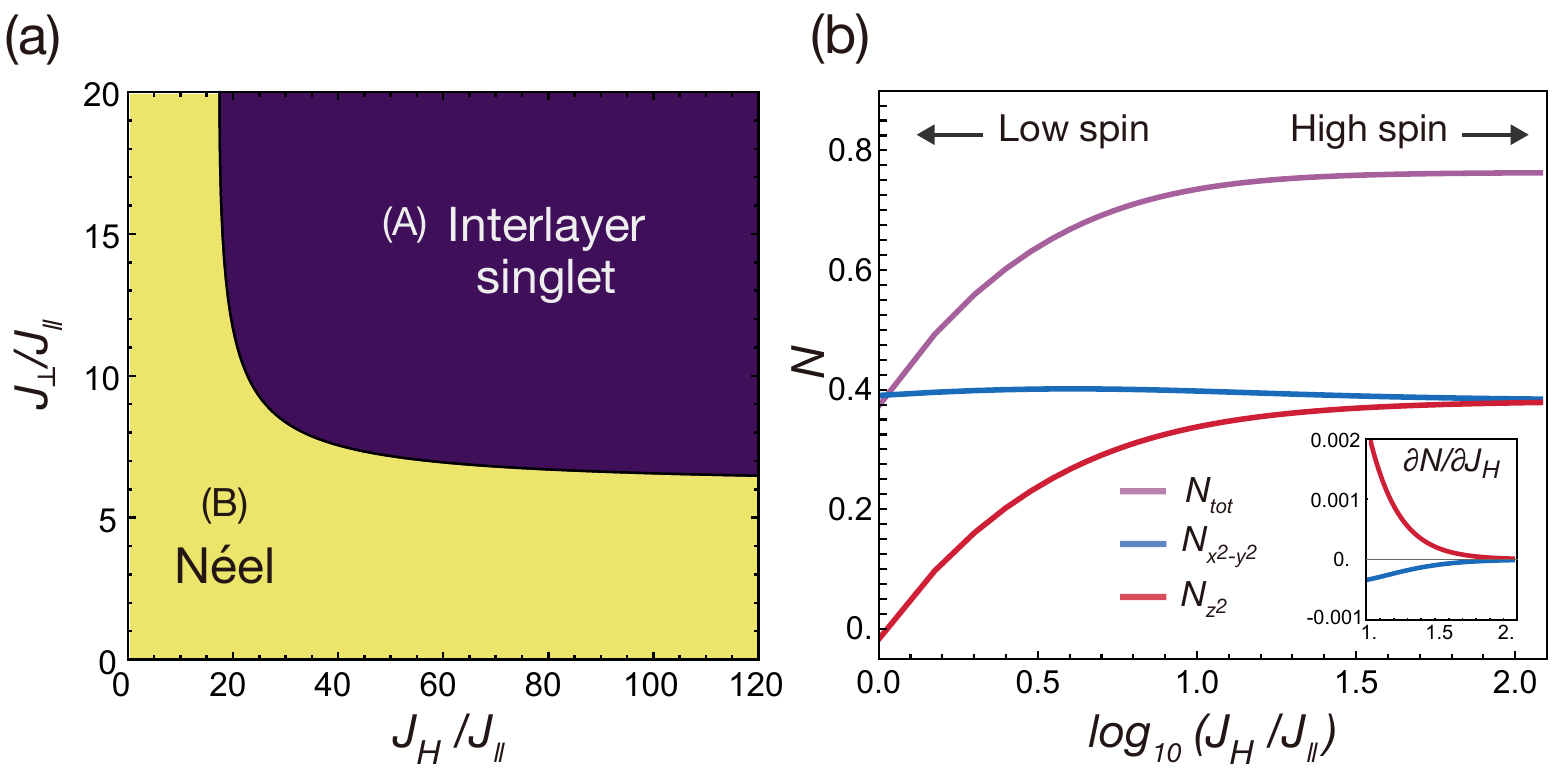}
    \caption{(a) Phase diagram from bond-operator mean-field theory (BOMFT). Blue, yellow region denotes interlayer singlet and N\'eel order, respectively. The black line indicates the second order phase transition boundary between these two phases.
(b) $J_H$ dependence of orbital-summed N\'eel order, $N=N_{x^2-y^2}+N_{z^2}$ (purple line) from Holstein-Primakov theory at $J_\perp = 5$. Orbital-resolved N\'eel orders, $N_{x^2-y^2} = (-1)^{i + l} \langle S^{d;z}_{i,l} \rangle$ and $N_{z^2} = (-1)^{i + l} \langle S^{f;z}_{i,l} \rangle$, are plotted in blue, red dashed line, respectively. The system transitions from low-spin (spin-1/2) AFM on the $X$-orbital at small $J_H$ to high-spin (spin-1) AFM from both orbitals at large $J_H$. 
Inset plots $\partial N / \partial J_H$ which highlights that $N_{x^2-y^2}$ decreases while $N_{z^2}$ increases with increasing $J_H$ at large $J_H$ regime.
}
    \label{fig:3}
\end{figure}

Within the BOMFT, one begins by assuming an interlayer singlet ground state and describes the system using a local singlet–triplet basis, which corresponds to the ground and first excited states in the $J_\parallel \rightarrow 0$ limit. 
The critical interlayer coupling $J_\perp^c$ is then determined by the point at which the triplet excitation gap closes, due to the finite in-plane coupling $J_\parallel$ effects.
Using two variational parameters—the condensation of interlayer singlet $\langle s \rangle$ and the chemical potential $\mu$—we numerically solve the self-consistent Hamiltonian and determine the phase boundary and its analytic form~\cite{SM}. 
Fig.~\ref{fig:3}(a) shows the resulting phase diagram from BOMFT. As expected, $J_\perp^c=\infty$ at $J_H=0$ and it decreases with increasing $J_H$ and approaches $J_\perp^c \rightarrow 6.10$ in the $J_H \rightarrow \infty$ limit, in closer agreement with the QMC result $J_\perp^c = 7.15$ for the bilayer spin-1 Heisenberg model~\cite{PhysRevB.84.214412}.
Therefore, the overall range of $J_\perp^{c}$ estimated by BOMFT is 
$6.10 < J_{\perp,c}^{\mathrm{BOMFT}} < \infty$. 
In contrast, the critical value obtained from the Schwinger-boson mean-field theory (SBMFT) 
lies in the range $13.5 < J_{\perp,c}^{\mathrm{SBMFT}} < \infty$, 
which is significantly larger than the BOMFT estimate. 
This discrepancy reflects the well-known tendency of mean-field approaches 
to overestimate the stability of ordered phases. 
Specifically, BOMFT (SBMFT) starts from the interlayer-singlet (N\'eel-ordered) limit, 
and therefore tends to underestimate (overestimate) $J_{\perp,c}$. 
Consequently, the true critical value should lie between the two mean-field estimates, 
$J_{\perp,c}^{\mathrm{BOMFT}} < J_{\perp,c} < J_{\perp,c}^{\mathrm{SBMFT}}$. 
Moreover, in the large-$J_H$ regime, the BOMFT results are expected to be closer to the actual value, as tabulated in Table.~S1, 
and they more faithfully capture the physics of the $J_\parallel$–$J_\perp$–$J_H$ model.

Next, to understand the spin-wave spectrum deep inside the antiferromagnetic phases, we use the standard Holstein–Primakov (HP) spin-wave theory~\cite{auerbach2012interacting,PhysRevB.92.245137}. 
We evaluate the orbital-resolved N\'eel order $N_{x^2-y^2} = (-1)^{i + l} \langle S^{d;z}_{i,l} \rangle$ and $N_{z^2} = (-1)^{i + l} \langle S^{f;z}_{i,l} \rangle$ and their sum, $N \equiv N_{x^2-y^2} + N_{z^2}$.
In Fig.~\ref{fig:3}(b), we plot $J_H$ dependence of $N$, $N_{x^2-y^2}$ and $N_{z^2}$ at a representative value $J_\perp = 5$, where the AFM phase is expected in all range of $J_H$.
For small $J_H$, $N_{z^2}$ remains suppressed and only $N_{x^2-y^2}$ exhibits finite magnetic order. 
This corresponds to a low-spin (spin-1/2) antiferromagnetic state. 
As $J_H$ increases, Hund's coupling more transmits $J_\perp$ to the $X$-orbital, suppressing $N_{x^2-y^2}$, while simultaneously enhancing the transmission of $J_\parallel$ to the $Z$-orbital, thereby increasing $N_{z^2}$ and $N$.
 In the large-$J_H$ limit, the two order parameters converge to the same value, indicating a high-spin (spin-1) antiferromagnetic state. Even in the $J_H \to \infty$ limit with a small $J_\perp$, $N\to 0.80$, matching the $S = 1$ HP result. 
 In SM, we present a full map of the order parameters across the $(J_\perp, J_H)$ space~\cite{SM}. 
By varying $J_\perp$, we find that the energy scale of $J_H$ associated with the crossover from the low-spin to the high-spin state decreases as $J_\perp$ is reduced.
 Thus, we have used several complementary approaches to establish that magnetism in the $d^8$ insulating regime can take on either high-spin, low spin forms, or can exhibit a spin-gapped, non magnetic ground state.

\textit{Doping toward the $d^{7.5}$ regime ---}
Having studied magnetism in the parent $d^8$ configuration, we now consider the effects of hole doping from the insulating $d^8$ Mott state. 
To qualitatively compare the doped high- and low-spin states, we study Eq.~\eqref{eq:two_hubbard} within the Hartree-Fock (HF) approximation.
We only investigate the robustness of the AFM orders under doping, and we will not consider the doped spin-gapped state here, as it has been studied elsewhere~\cite{yang2024strong,oh2024hightemperaturesuperconductivitykineticenergy}.
We decouple the interactions in Eq.~\eqref{eq:two_hubbard} via two auxiliary fields, $\vec{N}$ and $d\vec{N}$,
\begin{equation}
    \begin{aligned}
        H_\text{MF}=
       & H_0+\sum_{i\ell}({a_+\vec{N}^2}+{a_- d\vec{N}^2})-
        2(-1)^{i+\ell}\times\\
       &
        \left[(a_+\vec{N}+a_-d\vec{N})\cdot \vec{S}^d_{i\ell}+(a_+\vec{N}-a_-d\vec{N})\cdot \vec{S}^f_{i\ell}
        \right]
    \end{aligned}
\end{equation}
where $a_\pm=U/3\pm J_H/4$. Here, $\vec{N}=(-1)^{i+\ell}\langle \vec{S}^{d}_{i\ell}+\vec{S}^{f}_{i\ell}\rangle$ is an orbital-summed AFM order and $d\vec{N}=(-1)^{i+\ell}\langle \vec{S}^{d}_{i\ell}-\vec{S}^{f}_{i\ell}\rangle$ is the difference of two orbital-resolved AFM orders, $\vec{N}_{x^2-y^2}$ and $\vec{N}_{z^2}$. $H_0$ is non-interacting part of Eq.~\eqref{eq:two_hubbard}, which includes $t_\parallel,t_{\perp}, \Delta$. 
In the low spin AFM case where $Z$-orbitals tend to form interlayer singlets, we approximate the $Z$-orbitals as being in a disordered state, which corresponds to taking $a_+\vec{N}=a_-d\vec{N}$, such that the mean field terms do not couple to $\vec{S}^f_{i\ell}$. In the high-spin AFM, we take $d\vec{N}=0$ such that $\langle \vec{S}^d_{i\ell}\rangle  = \langle \vec{S}^f_{i\ell} \rangle$. In both cases, we consider orbital-summed N\'eel order $N$, and determine its expectation value self-consistently (see SM for details~\cite{SM}). We assume that the crystal field splitting $\Delta$ in Eq.~\eqref{eq:two_hubbard} is sufficiently large, so that the extra holes $x$ only enters to the $X$-orbitals. 
\begin{figure}
\includegraphics[width=0.95\linewidth]{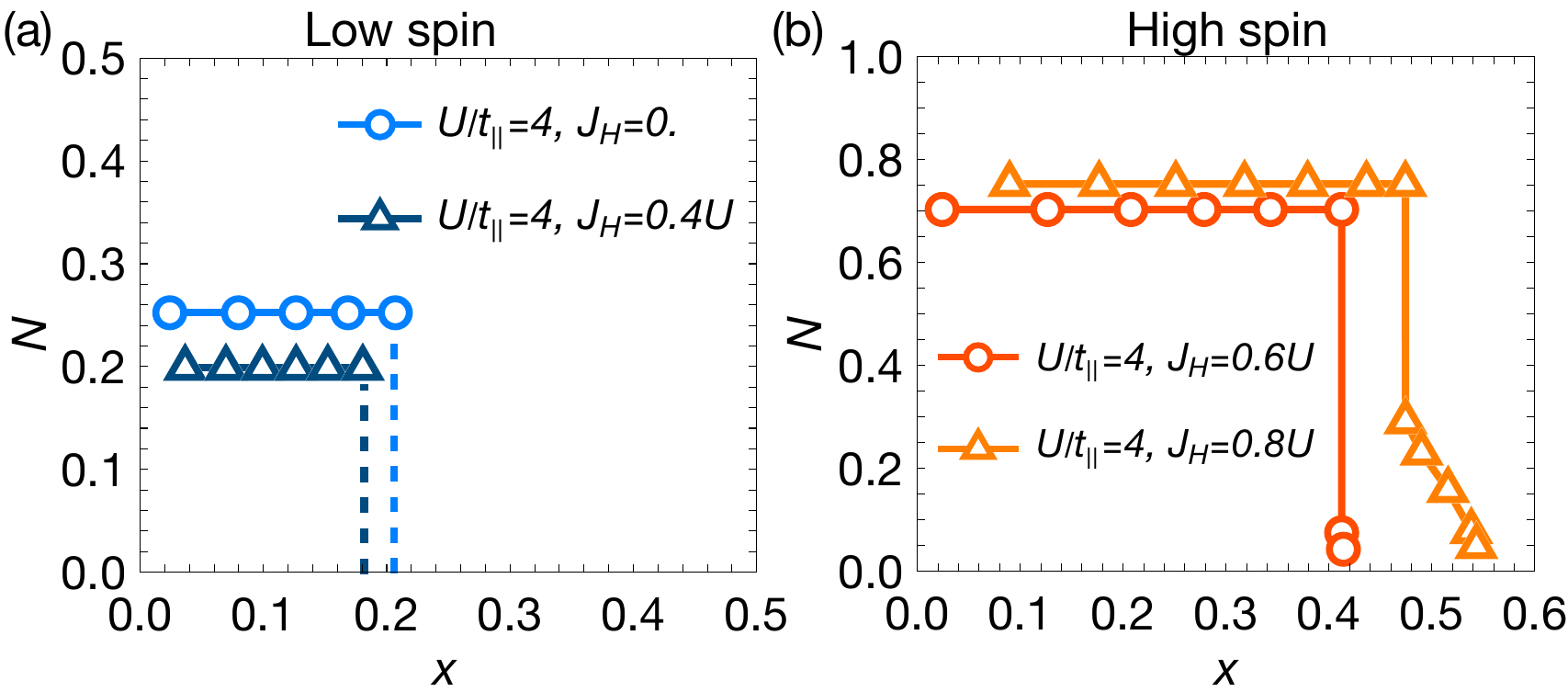}
    \caption{Hartree-Fock (HF) results for N\'eel order ($N$) under hole doping $x$ of (a) low-spin and (b) high-spin configurations for values of $U/t_\parallel$. 
    Here, $N=(-1)^{i+\ell}\langle S^{d;z}_{i,\ell}+S^{f;z}_{i,\ell}\rangle$ is an orbital-summed N\'eel order. 
    We used the hopping parameters $t_\parallel = 0.483$, $ t_\perp = 0.635$, and onsite energy difference $ \Delta = 0.367$ from DFT calculations~\cite{PhysRevLett.131.126001}. In both cases, we observe first-order or weakly first-order transitions. The AFM order in the high-spin case is more robust than in the low-spin case, as indicated by a larger critical doping $x_c $
}\label{fig:HFnumerical}
\end{figure}

We present the HF results for an intermediate $U=4t_\Vert$ in Fig.~\ref{fig:HFnumerical}. In the low spin case (Fig.~\ref{fig:HFnumerical}(a)), when the two orbitals are completely decoupled ($J_H=0$), the HF result coincides with the single layer single orbital square lattice Hubbard model, where for $U=4t$, AFM order vanishes at $x_c\approx 0.207$ through a first order phase transition~\cite{kivelson1994electronic}. 
Such first order antiferromagnetic phase transitions at $T=0$ are typical within the HF approximation. 
With a larger $J_H=0.4U$, the low spin AFM order gets suppressed compared to the decoupled limit: the spin order survives to a smaller $x_c\approx 0.18$ and its magnitude also becomes smaller.  By contrast, the high spin AFM can survive to a much larger hole doping, as can be seen in Fig.~\ref{fig:HFnumerical}(b). Moreover, the nature of the high spin AFM phase transition as a function of $x$ is also different from the low spin counterpart. We found that at large enough $J_H$, the high spin AFM order parameter $N$ can exhibit a discontinuous jump to a smaller value, which is due to the fact that the ground state energy has two local minima as a function of $N$, and increasing $x$ shifts the global minimum from one of two local minima to the other (see SM for details). But after this it appears to continuously decrease as $x$ increases further. This behavior seems to be a combination between both a first order and a continuous phase transition. Ref.~\cite{Xue_2024}, which employed a renormalized mean-field (Gutzwiller) approach starting from the type-II t-J model asuming infinite-Hund’s-coupling limit, reported a critical doping of $x_c \sim 0.55$ for the disappearance of antiferromagnetism (G-AFM).
This value is consistent with our result high spin results, despite the different model assumptions.

The conclusions based on the HF analysis are simple. The high-spin AFM order is more robust to doping than the low-spin counterpart. While the critical $x$ where magnetism is lost depends on the strength of $U$, we have found that for fixed $U$, magnetism in the high spin state survives out to higher hole concentration $x$. 
By contrast, when the $Z$-orbital spins are gapped out due to interlayer singlet formation, the resulting low spin state can resemble the overdoped cuprates. Due to crystal field effects, doped holes enter the $X$-orbital, and the $Z$-orbital remains half-filled.  As a consequence, the interlayer singlet formation by the $Z$-orbitals effectively removes these degrees of freedom leaving behind holes in a single band system. It is noteworthy that the bilayer geometry plays a crucial role here in allowing for interlayer singlet formation of the $Z$ spins.  In this case, the $d^{7.5}$ configuration corresponds to a hole concentration of $x=0.5$.  This value can be lowered further when one includes hybridization between the $X,Z$ orbitals, which results in charge transfer from $Z$ to $X$ orbitals.

Finally, we comment on some limitations of our Hartree–Fock (HF) calculation. 
In this work, we have considered only the simple N\'eel order, which is naturally expected to dominate near the $d^8$ parent compound. 
However, closer to the $d^{7.5}$ filling relevant to bilayer nickelates, the N\'eel order may compete with other spin-density-wave (SDW) configurations such as $(\pi/2,\pi/2)$ or $(\pi,0)$ orders~\cite{wang2025originspinstripesbilayer}, which should be included in a more comprehensive mean-field ansatz. 
Another open issue is possible phase separation, which has been extensively discussed in the single-layer $t$–$J$ model~\cite{PhysRevLett.64.475}. 
Investigating phase separation and other competing orders in the bilayer system remains an interesting direction for future work.

\textit{Discussion ---}
We have analyzed parent compounds of bilayer nickelate with $d^8$ electron configuration and have shown that there are several distinct types of magnetic behavior depending on the interplay between Hund's coupling ($J_H$) and superexchange ($J_\bot, J_\Vert$). We have identified a crossover from a low-spin AFM order at small $J_H$ to a high-spin AFM order at larger $J_H$.  We have also identified a spin-gapped insulating state that occurs when $J_\perp, J_H \gg J_\parallel$.
An important remark concerns the parameter regime $J_\perp, J_H \gg J_\parallel$ relevant to realistic materials.
While some DMFT-based analyses have estimated $J_H/J_\parallel \approx 5$ and $J_\perp/J_\parallel \sim 2$ for bilayer nickelates La$_3$Ni$_2$O$_7$ ($d^{7.5}$ configuration),
the exchange couplings are expected to evolve with doping $x$, and their realistic magnitudes in the $d^8$ limit remain an open question.

The nature of magnetism in the parent compound will certainly have implications for superconductivity.  We have shown here that the low energy physics of the doped low-spin state is adequately described by a single orbital bilayer Hubbard model when $J_H/J_\perp$ is vanishingly small. When $J_H/J_\perp$ is small, but finite, an exchange coupling between the two layers will be generated by virtual processes.  In contrast, the low energy effective description of the doped high spin state is a bilayer-Kondo lattice  with ferromagnetic Kondo coupling due to Hund's coupling and a weak interlayer exchange coupling. If one considers the extreme limit where $J_H \rightarrow \infty$, the Kondo model can be further simplified to a bilayer type-II $t$–$J$ model, as two of us derived in the previous paper~\cite{PhysRevB.108.174511}. We stress that all three different regimes can be reconciled by a double-Kondo model ~\cite{oh2024hightemperaturesuperconductivitykineticenergy}.

Within Hartree-Fock mean-field theory, we find that upon hole doping, 
the high-spin antiferromagnetic state becomes more robust. 
From this observation, it remains unclear how such robust magnetism 
influences superconductivity. 
Two opposite scenarios can be made: 
in one case, the enhanced magnetic order competes more strongly 
with superconductivity and may suppress pairing; 
in the other, within a spin-fluctuation framework, 
stronger magnetism could instead provide a more effective pairing ‘glue’ 
than in the low-spin counterpart. 
Lastly, the spin gapped insulator, when doped exhibits superconducivity rather distinct than either the high- or low-spin states, with strong interalyer pairing as was shown recently~\cite{yang2024strong,oh2024hightemperaturesuperconductivitykineticenergy}.  

Such considerations of high- and low-spin states are relevant to the broader family of nickel-based superconductors.  For example, the monolayer insulator La$_2$NiO$_4$ is known to be a spin-1 insulator~\cite{PhysRevLett.118.156402}.  Whether the bilayer remains in such a high spin state remains unclear; we have shown that a bilayer can well favor a low spin configuration via interlayer singlet formation of $Z$ spins.   Thus, a direct observation of the spin state of the bilayer would shed significant light on the nature of magnetism and superconductivity in these materials.  

Our findings are pertinent to recent experimental studies with compressively strained thin films of the bilayer nickelate. In particular, we encourage future experiments to explore the $d^8$ insulating limit and to probe possible magnetic transitions via strain engineering, which can be effectively described by varying $J_\perp/J_\parallel$ at fixed $J_H$.
 With compressive strain, we naturally expect that the in-plane superexchange will be enhanced whereas with tensile strain, the opposite would hold.  
Given the rich magnetic phase diagram revealed in our study, the experimental observation of distinct spin states in strain-tuned systems represents a promising arena for future exploration.

\textit{Acknowledgements ---}
S.R. would like to thank C. Varma for helpful discussions. The authors thank S. Kivelson for discussion abut the phase separation. We are also grateful to J. Li, B.Y. Wang and T. Helbig for useful discussions. 
H.O. and Y-H.Z. are supported by a startup fund from Johns Hopkins University.
Y.W., Y.Y., H.H., and S.R.  are supported in part by the US Department of Energy, Office of Basic Energy Sciences,
Division of Materials Sciences and Engineering, under Contract No. DE-AC02-76SF00515.
Y.W. acknowledges support from the Gordon and Betty Moore Foundation’s EPiQS Initiative through GBMF8686.
J.M.M. is supported by a startup fund at Stanford University.

%\bibliography{ref}

\onecolumngrid
\newpage
\clearpage
\setcounter{equation}{0}
\setcounter{figure}{0}
\setcounter{table}{0}
\setcounter{page}{1}
\setcounter{section}{0}

\maketitle 
\makeatletter
\renewcommand{\theequation}{S\arabic{equation}}
\renewcommand{\thefigure}{S\arabic{figure}}
\renewcommand{\thetable}{S\arabic{table}}
\renewcommand{\thesection}{S\arabic{section}}

\appendix
\onecolumngrid

\begin{center}
\vspace{10pt}
\textbf{\large Supplemental Material for ``High spin, low spin or gapped spins: minimal modeling \\
for the bilayer nickelates and magnetic parent compounds''}
\end{center} 
\begin{center} 
{Hanbit Oh$^{1,\textcolor{red}{*}}$, Yi-Ming Wu$^{2,\textcolor{red}{*}}$, Julian May-Mann$^{3}$, Yijun Yu$^{4,5}$,\\ Harold Y. Hwang$^{4,5}$, Ya-Hui Zhang$^{1,\textcolor{red}{\dagger}}$,
S. Raghu$^{2,4,\textcolor{red}{\ddagger}}$
}
\\

\emph{$^{1}$ William H. Miller III Department of Physics and Astronomy, \\
Johns Hopkins University, Baltimore, Maryland, 21218, USA}\\
\emph{$^{2}$ Stanford Institute for Theoretical Physics, Stanford University, Stanford, California 94305, USA}\\
\emph{$^{3}$ Department of Physics, Stanford University, Stanford, CA 94305, USA}\\
\emph{$^{4}$ Stanford Institute for Materials and Energy Sciences, SLAC National Accelerator Laboratory,
Menlo Park, CA 94025, USA}\\
\emph{$^{5}$ Departments of Applied Physics, Stanford University, Stanford, CA 94305, USA}
\vspace{5pt}
\end{center}
\tableofcontents

\section{I. Schwinger-Boson mean-field theory}

In the Schwinger-boson formalism, spin operators are represented by bosons $b$ as
\begin{equation}
\vec{S}^{a}_{i,l} = \frac{1}{2} \sum_{\sigma\sigma'} b^\dagger_{i,l,a,\sigma} \vec{\sigma}_{\sigma \sigma'} b_{i,l,a,\sigma'},
\end{equation}
subject to the local constraint
\begin{equation}
\sum_{\sigma} \langle b^\dagger_{i,l,a,\sigma} b_{i,l,a,\sigma} \rangle = 2S.
\label{eq:schwinger_number}
\end{equation}
Since the classical ground state of Eq.~(2) exhibits antiferromagnetic (AFM) order both within and between the layers, we divide the lattice into two sublattices: A ($i+l$ odd) and B ($i+l$ even). To encode AFM order, we apply a time-reversal transformation to sublattice B relative to A:
$b_{i,l,a,\uparrow} \rightarrow b_{i,l,a,\downarrow}, \quad 
b_{i,l,a,\downarrow} \rightarrow -b_{i,l,a,\uparrow}.
$
As a result, the spin operators are given by:
\begin{eqnarray}
(i,l) \in A: && 
S^{a,\mu}_{i,l} = \frac{1}{2} \sum_{\sigma\sigma'} b^\dagger_{i,l,a,\sigma} \sigma^\mu_{\sigma \sigma'} b_{i,l,a,\sigma'}, \label{eq:schwinger1} \\
(i,l) \in B: && 
S^{a,\mu}_{i,l} = \frac{1}{2} \sum_{\sigma\sigma'} b^\dagger_{i,l,a,\sigma} \tilde{\sigma}^\mu_{\sigma \sigma'} b_{i,l,a,\sigma'},
\label{eq:schwinger2}
\end{eqnarray}
where $\tilde{\sigma}^\mu = (-\sigma^x, \sigma^y, -\sigma^z)$ accounts for the time-reversal operation on sublattice B. Here, $l = 1,2$ labels the layer, $a = d,f$ denotes the $d_{x^2-y^2}$ and $d_{z^2}$ orbitals, and $\sigma = \uparrow,\downarrow$ is the spin index.

Substituting Eqs.~(\ref{eq:schwinger1},~\ref{eq:schwinger2}) into Eq.~(2), the Hamiltonian becomes
\begin{eqnarray}
H &=&
-\frac{J_\parallel}{2} \sum_{\langle i,j \rangle, l} 
\sum_{\sigma\sigma'}
b_{i,l,d,\sigma}^\dagger b_{j,l,d,\sigma}^\dagger 
b_{j,l,d,\sigma'} b_{i,l,d,\sigma'}
- \frac{J_\perp}{2} \sum_i 
\sum_{\sigma\sigma'} 
b_{i,1,f,\sigma}^\dagger b_{i,2,f,\sigma}^\dagger 
b_{i,2,f,\sigma'} b_{i,1,f,\sigma'} \nonumber \\
&&
- \frac{J_H}{2} \sum_{i,l} 
\sum_{\sigma\sigma'}
b^\dagger_{i,l,d,\sigma} b_{i,l,f,\sigma}
b^\dagger_{i,l,f,\sigma'} b_{i,l,d,\sigma'}.
\end{eqnarray}
We introduce the following mean-field order parameters:
\begin{eqnarray}
\Delta_{\parallel} &=& -2 \sum_{\sigma} \langle b_{i,l,d,\sigma} b_{j,l,d,\sigma} \rangle, \quad
\Delta_{\perp} = -\frac{1}{2} \sum_{\sigma} \langle b_{i,1,f,\sigma} b_{i,2,f,\sigma} \rangle, \nonumber \\
\Phi &=& -\frac{1}{2} \sum_{\sigma} \langle b^\dagger_{i,l,d,\sigma} b_{i,l,f,\sigma} \rangle.
\label{eq:order}
\end{eqnarray}
The resulting mean-field Hamiltonian becomes:
\begin{eqnarray}
H_{MF} &=&
\frac{J_\parallel}{4} \sum_{\langle i,j \rangle,l,\sigma} \Delta_\parallel b_{i,l,d,\sigma}^\dagger b_{j,l,d,\sigma}^\dagger + \text{h.c.}
+ J_\perp \sum_{i,\sigma} \Delta_\perp b_{i,1,f,\sigma}^\dagger b_{i,2,f,\sigma}^\dagger + \text{h.c.} \nonumber \\
&&
+ J_H \sum_{i,l,\sigma} \Phi b_{i,l,d,\sigma}^\dagger b_{i,l,f,\sigma} + \text{h.c.}
+ \sum_{a=d,f} \mu_a \sum_{i,l,\sigma} b_{i,l,a,\sigma}^\dagger b_{i,l,a,\sigma},
\label{eq:ham_mf}
\end{eqnarray}
where $\mu_a$ are Lagrange multipliers that enforce the local constraint, Eq.~(\ref{eq:schwinger_number}).

After Fourier transformation, we 
diagonalize the Hamiltonian via paraunitary transformation yielding, 
\begin{equation}
H_{MF} = \sum_{k,\sigma} \sum_{A=1}^4 \Omega_{k,A} \, \tilde{\psi}^\dagger_{k,A,\sigma} \tilde{\psi}_{k,A,\sigma},
\end{equation}
where the bosonic dispersions are
\begin{eqnarray}
\{\Omega_{k,1}, \Omega_{k,2}, \Omega_{k,3}, \Omega_{k,4}\}
= \left\{
\frac{1}{\sqrt{2}} \sqrt{X_k \pm \sqrt{X_k^2 - 4Y_k}}, \;
\frac{1}{\sqrt{2}} \sqrt{X_{k+Q} \pm \sqrt{X_{k+Q}^2 - 4Y_{k+Q}}}
\right\},
\end{eqnarray}
with $\vec{Q} = (\pi,\pi)$ and
\begin{eqnarray}
X_k &=& \mu_d^2 + \mu_f^2 - (J_\parallel \Delta_\parallel \gamma_k)^2 - (J_\perp \Delta_\perp)^2 + 2 (J_H \Phi)^2, \\
Y_k &=& \left[ (\mu_d + J_\parallel \Delta_\parallel \gamma_k)(\mu_f + J_\perp \Delta_\perp) - (J_H \Phi)^2 \right] 
\left[ (\mu_d - J_\parallel \Delta_\parallel \gamma_k)(\mu_f - J_\perp \Delta_\perp) - (J_H \Phi)^2 \right],
\end{eqnarray}
where $\gamma_k = (\cos k_x + \cos k_y)/2$.

The condition for real, stable bosonic modes (i.e., a positive-definite Hamiltonian) requires
\begin{eqnarray}
(\mu_d - J_\parallel \Delta_\parallel) + (\mu_f - J_\perp \Delta_\perp) 
> \sqrt{
\left[ (\mu_d - J_\parallel \Delta_\parallel) - (\mu_f - J_\perp \Delta_\perp) \right]^2 + 4(J_H \Phi)^2
}.
\end{eqnarray}
We numerically solve the self-consistent equations, Eqs.~(\ref{eq:order})–(\ref{eq:ham_mf}), by iteratively updating the variational parameters $(\Delta_\parallel, \Delta_\perp, \Phi)$. At each step, $\mu_d$ and $\mu_f$ are adjusted to enforce the constraint Eq.~(\ref{eq:schwinger_number}). Stability condition is ensured by satisfying the following inequalities:
\begin{eqnarray}
\mu_d &> J_\parallel \Delta_\parallel + \frac{(J_H \Phi)^2}{\mu_f - J_\perp \Delta_\perp}, 
\quad \text{with } \mu_f > J_\perp \Delta_\perp, \\
\mu_f &> J_\perp \Delta_\perp + \frac{(J_H \Phi)^2}{\mu_d - J_\parallel \Delta_\parallel}, 
\quad \text{with } \mu_d > J_\parallel \Delta_\parallel.
\end{eqnarray}

In our numerical implementation, we approximate the momentum summation $\sum_k = \int d\gamma, D(\gamma)$ by treating the density of states $D(\gamma)$ as a constant value of $1/2$, following the standard approximation used in prior literature~\cite{PhysRevB.53.12196}. For numerical integration over $\gamma$, we discretize the interval $\gamma \in [-1, 1]$ using a step size of $1/2000$.

\begin{figure}
    \centering
\includegraphics[width=0.35\linewidth]{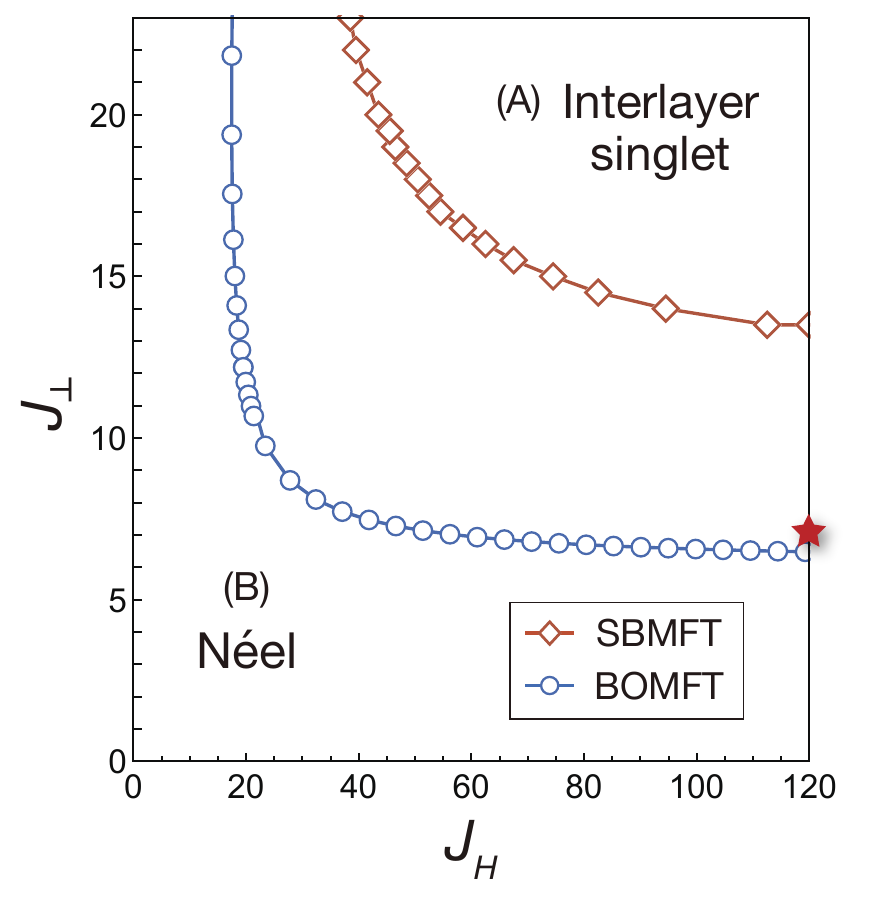}
    \caption{Phase diagram from Schwinger-boson mean-field theory and bond-operator mean-field theory.
Red and blue lines denote the critical coupling $J_\perp^{c}$ separating the N\'eel and disordered phases, obtained from Schwinger-boson mean-field theory (SBMFT) and bond-operator mean-field theory (BOMFT), respectively. The red star indicates the QMC result of $J_\perp^c=7.15$ for the spin-1 bilayer Heisenberg model~\cite{PhysRevB.84.214412}.
    }
    \label{fig:s0}
\end{figure}

\begin{table}[h]
    \centering
    \begin{tabular}{c|c|c|c|c|c}
    \hline
    \hline
      Method&   SB-MFT&BO-MFT& BO-MFT+triplet &   
     QMC of bilayer $S=1$ \\
         \hline
        $J_{\perp,c}[J_H\rightarrow\infty]$
        &
        11 & 
        6.09 &
        6.37 &
        7.150(2) \cite{PhysRevB.84.214412}
        \\
    \hline
    \hline
    \end{tabular}
    \caption{\textbf{Comparison of $J_{\perp,c}$ of our model, Eq.(2) with various methods at $J_H\rightarrow \infty$ limit. }
    }
    \label{table1}
\end{table}

\section{II. Bond-operator mean-field theory}
In this section, we study N\'eel to interlayer singlet transition using the different mean-field theories following the bond-operator representation \cite{PhysRevB.84.214412}.  

\subsection{1. Bond operators}

\begin{figure}[h]
    \centering
\includegraphics[width=.65\linewidth]{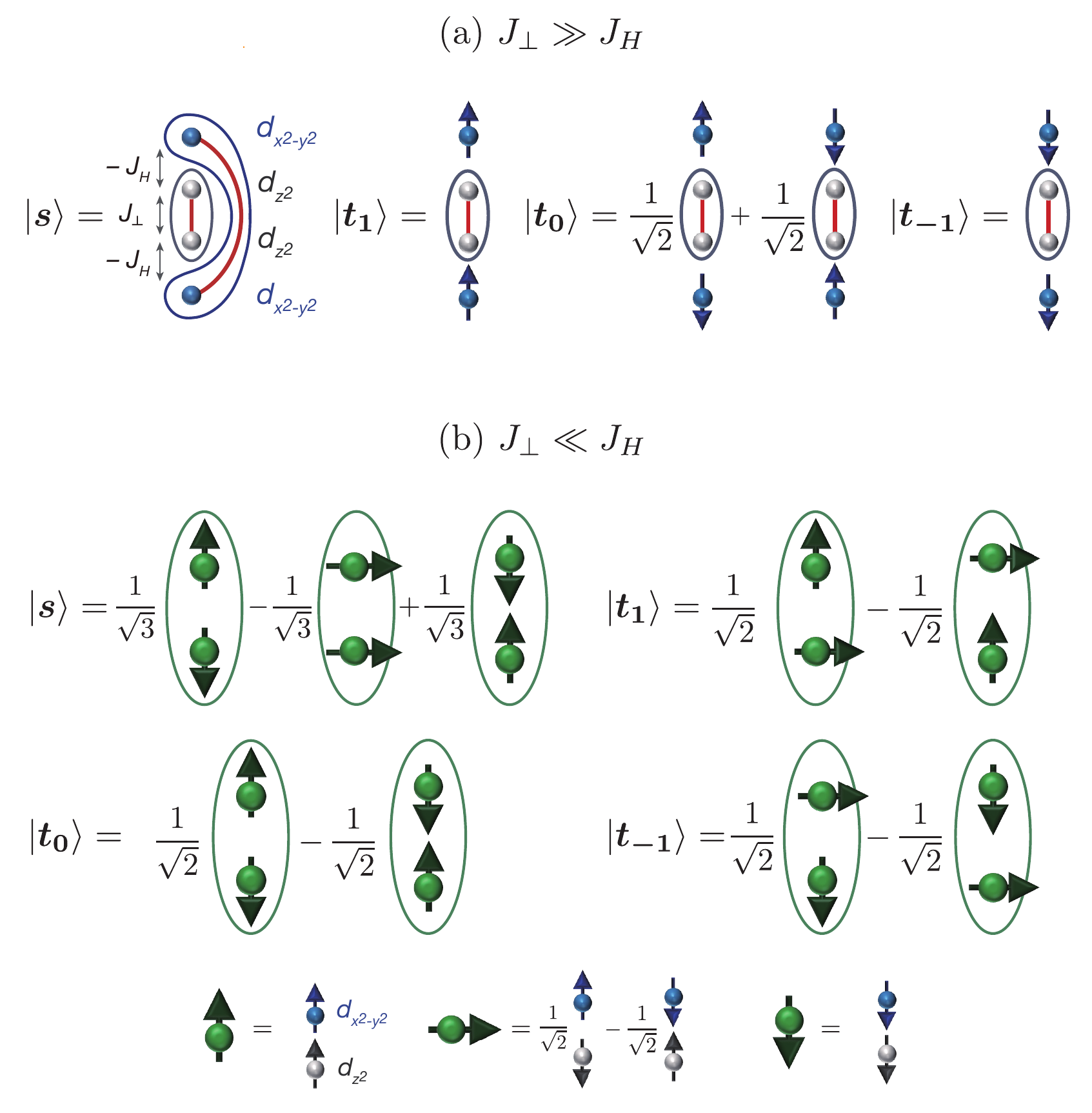}
\caption{Illustration of the interlayer singlet $|s\rangle$ and interlayer triplet $|t_m\rangle$ states, depending on the ratio $J_H/J_\perp$.
We illustrate two representative cases: (a) $J_\perp \gg J_H$ and (b) $J_\perp \ll J_H$. 
Two blue and white spheres represent the $d_{x^2 - y^2}$ and $d_{z^2}$ orbitals from each layers, each carrying a spin-$1/2$ moment. 
In (a), black ellipses with red lines highlight the interlayer singlet formed by two $d_{z^2}$ orbitals across two layers. 
In (b), green arrows indicate spin-1 moments formed via ferromagnetic alignment between $d_{x^2 - y^2}$ and $d_{z^2}$ spins, as illustrated in the last lines.}
    \label{fig:s1}
\end{figure}

\begin{figure}[tb]
    \centering
\includegraphics[width=.9\linewidth]{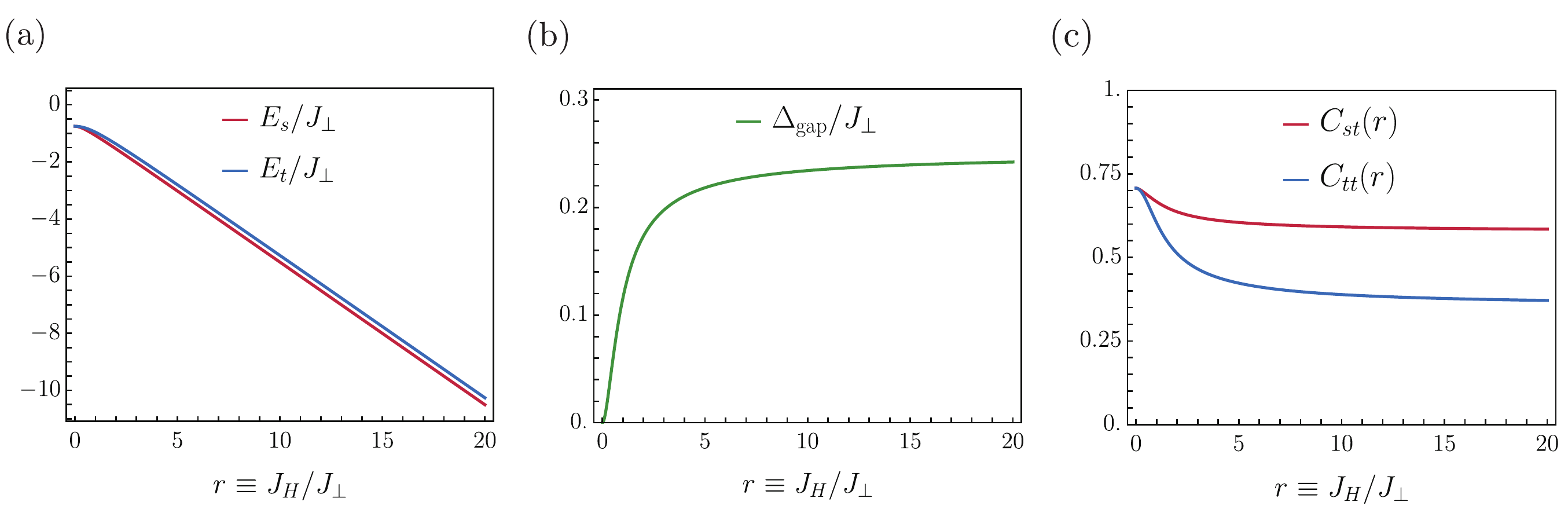}
\caption{$r=J_H/J_\perp$ dependence of (a) the energy of interlayer singlet $E_{s}(r)$ and triplet $E_t(r)$, (b) their energy gap, $\Delta_{\mathrm{gap}}(r)=E_t(r)-E_s(r)$, and (c) $C_{st}(r)$, $C_{tt}(r)$ in Eqs.~(\ref{eq:S_plus},~\ref{eq:S_z}).
}
    \label{fig:s2}
\end{figure}

In the $J_\parallel \rightarrow 0$ limit, we can define local singlet and triplet states. In our model, each site contains four $S=1/2$ moments—two orbitals on each of the two layers—which decompose into irreducible spin multiplets as
$\bm{1/2}\otimes\bm{1/2}\otimes\bm{1/2}\otimes\bm{1/2} =
2 \times \bm{0}
\oplus
3 \times \bm{1}
\oplus
1 \times \bm{2}
$.
That is, the local Hilbert space consists of two singlets, nine triplet states, and five quintet states. While the detailed structure of these multiplets depends on the ratio $r = J_H / J_\perp$, for all values of $r$, the lowest-energy state is a singlet, denoted by $\ket{s} \equiv s^\dagger \ket{0}$, and the first excited states are the three triplet modes $\ket{t_m} \equiv t^\dagger_m \ket{0}$ with $m = 1, 0, -1$. 
In Fig.~\ref{fig:s1}, we illustrate the form of the $\ket{s}$ and $\ket{t_m}$ for $J_\perp\gg J_H$ and $J_\perp \ll J_H$ limit.
In Fig.~\ref{fig:s2}(a,b), we plot the energy of the singlet and triplet states and their energy gap depending on $J_\perp/ J_H$.

We now incorporate a finite in-plane exchange $J_\parallel$ by expressing the spin operators of the $d_{x^2 - y^2}$ orbital in terms of the interlayer singlet and triplet basis. 
Within the restricted Hilbert space spanned by $\{\ket{s}, \ket{t_{m}}\}$, the spin operators for the $d_{x^2 - y^2}$ orbital ($\vec{S}^d$) is written in the form:
\begin{eqnarray}
S_{i,l}^{d,+} &=& (-1)^l \,
    C_{st}(r)\,  \left(
    s_i^\dagger t_{i,-1} - t^{\dagger}_{i,+1} s_i\right)
    + C_{tt}(r)\left(t^{\dagger}_{i,1} t_{i,0} + t^{\dagger}_{i,0} t_{i,-1}\right),
\label{eq:S_plus} \\
S^{d,z}_{i,l} &=& (-1)^l \,
    \frac{C_{st}(r)}{\sqrt{2}}\,  \left(s_{i}^\dagger t_{i,0} + t^{\dagger}_{i,0} s_{i}\right)
    + \frac{C_{tt}(r)}{\sqrt{2}} \left(t^{\dagger}_{i,1} t_{i,1} - t^{\dagger}_{i,-1} t_{i,-1}\right),
\label{eq:S_z}
\end{eqnarray}
where $C_{st}(r)$ and $C_{tt}(r)$ are dimensionless coefficients that depend on the ratio $r = J_H / J_\perp$, and reflect the overlap between singlet and triplet wavefunctions. The indices $i$ and $l$ denote the rung site and layer ($l = 1,2$ for top and bottom), respectively. The $r$-dependence of $C_{st}(r)$ and $C_{tt}(r)$ is shown in Fig.~\ref{fig:s2}(c).

\subsection{2. Singlet-triplet mean-field theory}

We now rewrite the Hamiltonian in Eq.~(\ref{eq:two_heisenberg}) in the singlet–triplet basis, which consists of two parts: (i) the $J_\perp$ and $J_H$ terms contribute to onsite energies of the singlet and triplet states, and (ii) the in-plane coupling $J$ is expressed using the spin operators in Eqs.~(\ref{eq:S_plus})–(\ref{eq:S_z}). At the mean-field level, we assume uniform condensation of singlet bosons with $\overline{s} = \langle s_i \rangle$.
Under this assumption, the mean-field Hamiltonian takes the form:
\begin{eqnarray}
H_{MF} &=& E_s(r)\, \overline{s}^2 \mathcal{N} + E_t(r) \sum_{i,m} t^\dagger_{i,m} t_{i,m} 
+ \mu \left( \sum_{i,m} t^\dagger_{i,m} t_{i,m} + \overline{s}^2 \mathcal{N} - \mathcal{N} \right) \nonumber \\
&& + C_{st}^2(r)\, \overline{s}^2 \sum_{\langle i,j \rangle} \left[ (t_{i,0} + t^\dagger_{i,0})(t_{j,0} + t^\dagger_{j,0}) + (t_{i,-1} - t^\dagger_{i,1})(t_{j,-1} - t^\dagger_{j,1}) + \text{h.c.} \right],
\end{eqnarray}
where $\mu$ is a Lagrange multiplier enforcing the constraint $\sum_m t^\dagger_{i,m} t_{i,m} + \overline{s}^2 = 1$ at each site. $\mathcal{N}$ is a number of sites. 

Retaining only quadratic terms in the triplet operators, the mean-field Hamiltonian per site becomes:
\begin{eqnarray}
\frac{H_{MF}}{\mathcal{N}} &=& E_s\, \overline{s}^2 + \mu(\overline{s}^2 - 1) - \frac{3}{2} A \nonumber \\
&& + \frac{1}{\mathcal{N}} \sum_{\vec{k}, \alpha = x,y,z} 
\left( \begin{array}{cc} t^\dagger_{\vec{k},\alpha} & t_{-\vec{k},\alpha} \end{array} \right)
\left( \begin{array}{cc} A + 2\epsilon_{\vec{k}} & 2\epsilon_{\vec{k}} \\ 2\epsilon_{\vec{k}} & A + 2\epsilon_{\vec{k}} \end{array} \right)
\left( \begin{array}{c} t_{\vec{k},\alpha} \\ t^\dagger_{-\vec{k},\alpha} \end{array} \right),
\end{eqnarray}
where $A = E_t + \mu$ and $\epsilon_{\vec{k}} = C_{st}^2(r)\, \overline{s}^2 (\cos k_x + \cos k_y)$.
The ground-state energy per site is:
\begin{eqnarray}
\frac{E_0}{\mathcal{N}} =E_s\, \overline{s}^2 + \mu (\overline{s}^2 - 1) - \frac{3}{2} A + \frac{3}{\mathcal{N}} \sum_{\vec{k}} \lambda_{\vec{k}},
\end{eqnarray}
where $\lambda_{\vec{k}} = \sqrt{A(A + 4\epsilon_{\vec{k}})}$ is the triplon dispersion.

The mean-field solution is obtained by minimizing $E_0$ with respect to $\mu$ and $\overline{s}^2$:
\begin{eqnarray}
\overline{s}^2 &=& \frac{5}{2} - \frac{3}{2} \sum_{\vec{k}} \frac{A + 2\epsilon_{\vec{k}}}{\lambda_{\vec{k}}}, \\
-\mu &=& E_s + \frac{6}{\mathcal{N}} \sum_{\vec{k}} \frac{A \epsilon_{\vec{k}}}{\overline{s}^2 \lambda_{\vec{k}}}.
\end{eqnarray}

The critical interlayer coupling $J_{\perp,c}(r)$ is determined by the gap-closing condition $A + 4\epsilon_{\pi,\pi} = 0$, which yields:
\begin{eqnarray}
J_\perp^c(r) = \frac{C_{st}^2(r)}{E_t(r) - E_s(r)} 
\left[ 20 - \frac{24}{\mathcal{N}} \sum_{\vec{k}} \frac{1}{\sqrt{1 + \frac{1}{2}(\cos k_x + \cos k_y)}} \right],
\label{eq:critical_Jp}
\end{eqnarray}
where the bracketed term evaluates numerically as $4.57083$. The $r\equiv J_H/J_\perp$-dependent coefficients $C_{st}(r)$, $E_s(r)$, and $E_t(r)$ are plotted in Fig.~\ref{fig:s2}.

\section{III. Holstein–Primakoff spin-wave theory}
In this section, we present the details of our spin-wave calculations based on the standard Holstein–Primakoff (HP) formalism. In this approach, the quantum spin Hamiltonian is expressed using spin operators defined in the local basis of classical spin orientations. As in the Schwinger-boson treatment, the classical ground state of Eq.~(\ref{eq:two_heisenberg}) exhibits antiferromagnetic (AFM) order both within and between layers. Accordingly, we introduce two sublattices, A and B.

The spin operators on each sublattice are transformed as:
\begin{eqnarray}
\text{For } (i,l) \in A:\quad 
(S_{i,l}^{a,x},S_{i,l}^{a,y},S_{i,l}^{a,z})
&=&(\tilde{S}_{i,l}^{a,x},\tilde{S}_{i,l}^{a,y},\tilde{S}_{i,l}^{a,z}),
\label{eq:HP1} \\
\text{For } (i,l) \in B:\quad 
(S_{i,l}^{a,x},S_{i,l}^{a,y},S_{i,l}^{a,z})
&=&(-\tilde{S}_{i,l}^{a,x},\tilde{S}_{i,l}^{a,y},-\tilde{S}_{i,l}^{a,z}),
\label{eq:HP2}
\end{eqnarray}
where the Holstein–Primakoff representation is given by
\begin{eqnarray}
    \tilde{S}^{a,+}_{i,l} = \sqrt{2S - n_{b_{i,l,a}}} \, b_{i,l,a}, \quad
    \tilde{S}^{a,-}_{i,l} = b_{i,l,a}^\dagger \sqrt{2S - n_{b_{i,l,a}}}, \quad
    \tilde{S}^{a,z}_{i,l} = S - n_{b_{i,l,a}}.
    \label{eq:HP3}
\end{eqnarray}
Here, $l = 1,2$ denotes the layer, and $a = d,f$ labels the $d_{x^2 - y^2}$ and $d_{z^2}$ orbitals, respectively. The bosonic operators satisfy the canonical commutation relation $[b_{i,l,a}, b^\dagger_{i',l',a'}] = \delta_{i,i'} \delta_{l,l'} \delta_{a,a'}$.

Substituting Eqs.~(\ref{eq:HP1})–(\ref{eq:HP3}) into Eq.~(\ref{eq:two_heisenberg}) yields the expansion $H = H_0 + H_1 + \cdots$, where $H_0 = \mathcal{O}(S^2)$ is the classical ground-state energy and $H_1 = \mathcal{O}(S)$ provides the linear spin-wave spectrum. The explicit form of $H_1$ is
\begin{eqnarray*}
H_1 &=&
2JS \sum_{\vec{k},\eta=\pm 1} \left[
\gamma_{\vec{k}} \left( b_{\vec{k},\eta,d} b_{-\vec{k},\eta,d} + \text{h.c.} \right)
+ 2 b_{\vec{k},\eta,d}^\dagger b_{\vec{k},\eta,d}
\right] \nonumber\\
&& + \frac{J_\perp S}{2} \sum_{\vec{k},\eta=\pm 1} \left[
\eta \left( b_{\vec{k},\eta,f} b_{-\vec{k},\eta,f} + \text{h.c.} \right)
+ 2 b_{\vec{k},\eta,f}^\dagger b_{\vec{k},\eta,f}
\right] \nonumber\\
&& - J_H S \sum_{\vec{k},\eta=\pm 1} \left[
b_{\vec{k},\eta,d} b_{\vec{k},\eta,f}^\dagger 
+ b_{\vec{k},\eta,d}^\dagger b_{\vec{k},\eta,f}
- b_{\vec{k},\eta,d}^\dagger b_{\vec{k},\eta,d}
- b_{\vec{k},\eta,f}^\dagger b_{\vec{k},\eta,f}
\right],
\end{eqnarray*}
where $\gamma_{\vec{k}} = (\cos k_x + \cos k_y)/2$, and $\eta = \pm 1$ labels symmetric and antisymmetric combinations under layer exchange.

After applying a paraunitary transformation, the Hamiltonian is diagonalized as
\begin{eqnarray}
H_1 = S \sum_{\vec{k},\eta=\pm, a=1,2} \Omega_{\vec{k},\eta,a} \, \tilde{b}_{\vec{k},\eta,a}^\dagger \tilde{b}_{\vec{k},\eta,a} + \cdots,
\end{eqnarray}
with the spin-wave dispersions
\begin{eqnarray*}
\left\{ \Omega_{\vec{k},\eta,1}, \Omega_{\vec{k},\eta,2} \right\}
= \sqrt{f_0^2 + f_1^2 + f_3^2 - g_0^2 - g_3^2 
\pm 2 \sqrt{
f_0^2(f_1^2 + f_3^2) + g_0^2 g_3^2 - f_1^2 g_3^2 
- 2 f_0 f_3 g_0 g_3
}},
\end{eqnarray*}
where
\begin{eqnarray*}
f_0 = J + \frac{J_\perp}{4} + \frac{J_H}{2}, \quad
f_1 = -\frac{J_H}{2}, \quad
f_3 = J - \frac{J_\perp}{4}, \\
g_0(\vec{k},\eta) = J \gamma_{\vec{k}} + \frac{J_\perp}{4} \eta, \quad
g_3(\vec{k},\eta) = J \gamma_{\vec{k}} - \frac{J_\perp}{4} \eta.
\end{eqnarray*}

Analyzing $\Omega_{\vec{k},\eta,a}$, the bosonic gap closes only when
\begin{eqnarray}
\gamma_{\vec{k}} = \eta \quad \text{or} \quad \gamma_{\vec{k}} = \eta \left[1 + \frac{J_H J_\perp}{2J(J_H + 4J_\perp)}\right].
\end{eqnarray}
The first condition occurs at $\vec{k} = (0,0)$ and $(\pi,\pi)$, while the second condition is never satisfied for $J, J_\perp, J_H > 0$. Thus, gapless spin-wave excitations appear only at $(0,0)$ and $(\pi,\pi)$, independent of the parameter values. This suggests that linear spin-wave theory overestimates the extent of the AFM phase compared to other approaches.

\begin{figure}[h]
    \centering
\includegraphics[width=0.85\linewidth]{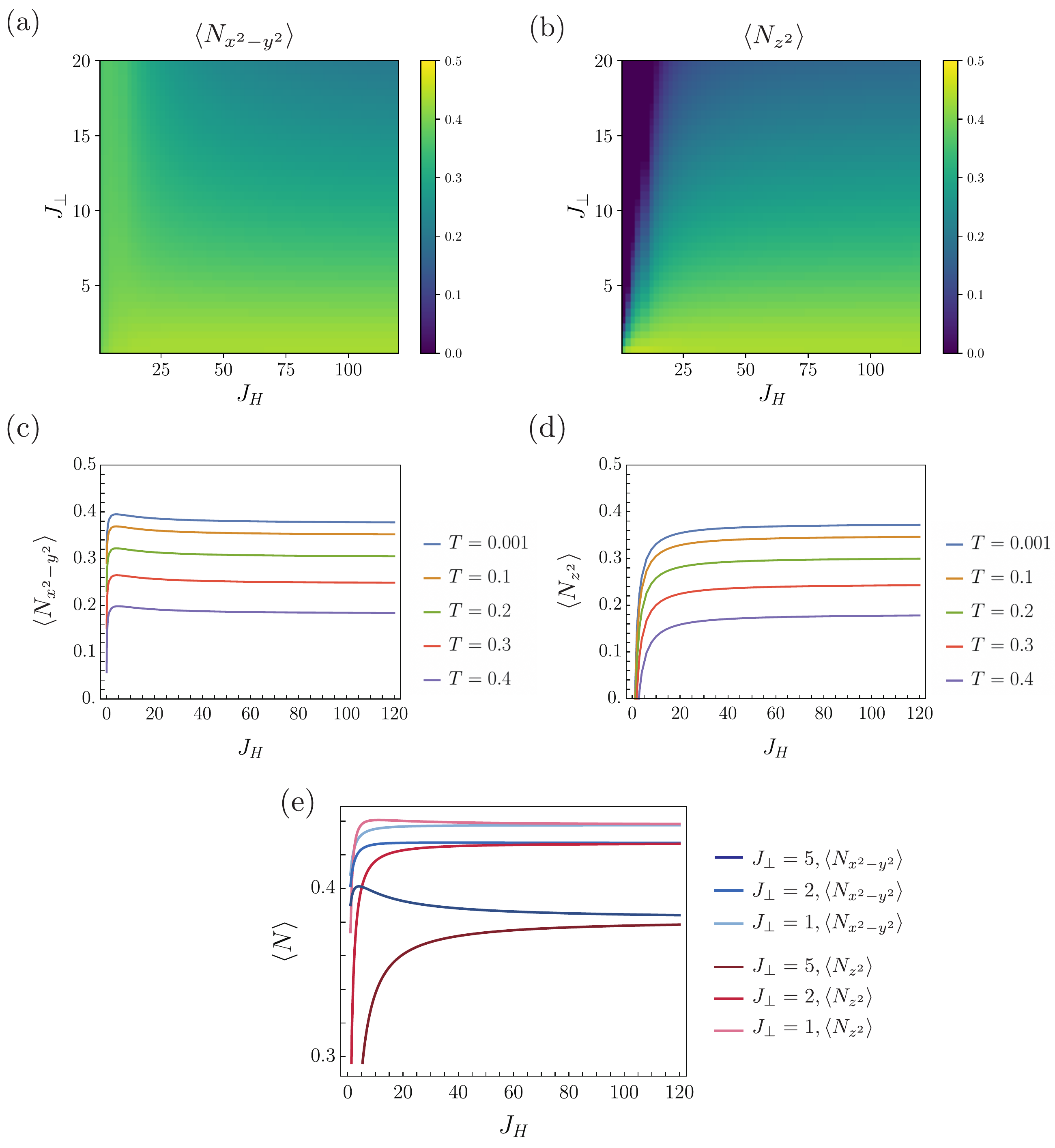}
    \caption{
   Holstein-Primakoff spin-wave theory results with $S=1/2$ for $\langle N_{x^2 - y^2} \rangle$ and $\langle N_{z^2} \rangle$. Here, we fix $J=1$.
(a,b) Zero-temperature $(J_H, J_\perp)$ dependence of the staggered moments $\langle N_{x^2 - y^2} \rangle$ and $\langle N_{z^2} \rangle$. The $d_{x^2 - y^2}$ moment grows with increasing $J_H$ and decreasing $J_\perp$, while the $d_{z^2}$ moment is enhanced for small $J_H$ and small $J_\perp$.
 As $J_H \to 0$, $\langle N_{x^2 - y^2} \rangle$ approaches 0.30, consistent with the known zero-temperature HP result for a square lattice with $S = 1/2$.
 As $J_H \to 0$, $\langle N_{x^2 - y^2} \rangle$ approaches 0.30, consistent with the HP result for a square lattice with $S = 1/2$. In the $J_H \to \infty$ and small $J_\perp$ limit, $\langle N_{x^2 - y^2} \rangle + \langle N_{z^2} \rangle \to 0.80$, matching the $S = 1$ result.
(c,d) Temperature dependence of $\langle N_{x^2 - y^2} \rangle$ and $\langle N_{z^2} \rangle$ at fixed $J_\perp = 5$.
(e) Order parameter at different $J_\perp$. This demonstrates that the energy scale of $J_H$ associated with the crossover from the low-spin to the high-spin state decreases as $J_\perp$ is reduced.
}
    \label{fig:s3}
\end{figure}

We also evaluate the N\'eel order parameter for each orbital $a = d (x^2-y^2), f(z^2)$, defined as
\begin{eqnarray}
N_a \equiv \frac{1}{2N} \sum_i \langle S_{i,1}^{a,z} - S_{i,2}^{a,z} \rangle = S - \frac{1}{2\mathcal{N}} \sum_{\vec{k},\eta=\pm} \langle b_{\vec{k},\eta,a}^\dagger b_{\vec{k},\eta,a} \rangle,
\label{eq:Na}
\end{eqnarray}
where temperature dependence enters through the Bose–Einstein distribution $n_{\mathrm{BE}}(\Omega_{\vec{k},\eta,a}/T) = \langle \tilde{b}_{\vec{k},\eta,a}^\dagger \tilde{b}_{\vec{k},\eta,a} \rangle$. $\mathcal{N}$ is number of the sites.
In Fig.~\ref{fig:s3}, we plot the $(J_H,J_\perp)$ and $T$ dependence of  $\langle N_{x^2 - y^2} \rangle$ and $\langle N_{z^2} \rangle$ by sustituting $S=1/2$ in Eq.~(\ref{eq:Na}). For numerical integration over $\vec{k}$, we discretize the interval $k_x,k_y\in [-\pi, \pi]$ using a step size of $2\pi/200$.

\section{IV. HF calculation of the low- and high-spin AFM under doping} % (fold)
\label{sec:hf_calculation_of_the_low_and_high_spin_afm_under_doping}

In this section, we present a mean field estimation of how the low-spin and high-spin AFM orders at $d^8$ configuration evolve under doping towards $d^{7.5}$ configuration. We note that for the low-spin AFM, the mean field treatment does not capture the rung-singlet state for the $d_{z^2}$ orbitals, therefore we approximate this state as spin disordered state.  

We consider this model
\begin{align}
H &= H_0 + U \sum_{i\ell} \left( n^d_{i\ell\uparrow} n^d_{i\ell\downarrow} + n^f_{i\ell\uparrow} n^f_{i\ell\downarrow} \right)
- J_H \sum_{i\ell} \vec{S}^d_{i\ell} \cdot \vec{S}^f_{i\ell},
\label{eq:two_hubbard2}
\end{align}
where the non-interacting part is 
\begin{equation}
    H_0=\sum_{\vec{k},\sigma}(d_{\vec{k}\sigma}^\dagger, f_{\vec{k}\sigma}^\dagger)\begin{pmatrix}
        \varepsilon^d(\vec{k}) & 0\\
        0 & \varepsilon^f(\vec{k})
    \end{pmatrix}\begin{pmatrix}
        d_{\vec{k}\sigma}\\
        f_{\vec{k}\sigma}
    \end{pmatrix}
\end{equation}
Here the dispersions are

\begin{equation}
     \begin{aligned}
         &\varepsilon^d(\vec{k})=-2t_\Vert(\cos k_x +\cos k_y)-\mu_d,\\
         &\varepsilon^f(\vec{k})=-t_\bot \cos k_z -\mu_f.\\
     \end{aligned}\label{eq:H0}
 \end{equation} 
and we note $k_z$ can only take two values $k_z=0$ or $k_z=\pi$, corresponding to bonding and anti-bonding bands, and there is no factor of 2 in front of $t_\bot\cos k_z$. 
We the parameters in Eq.\eqref{eq:H0} are taken from Ref., which we list below
\begin{equation}
    \begin{aligned}
         &t_\Vert=0.483,  ~~~ \mu_d=-0.776;\\
         &t_\bot=0.635,~~~ \mu_f=-0.409;
    \end{aligned}
\end{equation}

Using the identity
\begin{equation}
    n_{i\uparrow}n_{i\downarrow}=-\frac{2}{3}\vec{S}_i^2+\frac{1}{2}n_i
\end{equation}
for both orbitals
and absorbing the constants to the redefinition of chemical potential, we rewrite the Hamiltonian as 
\begin{equation}
    \begin{aligned}
        H &= H_0-\frac{2U}{3}\sum_{i\ell}\left((\vec{S}_{i\ell}^d)^2+(\vec{S}_{i\ell}^f)^2\right)-J_H \sum_{i\ell} \vec{S}^d_{i\ell} \cdot \vec{S}^f_{i\ell}\\
        &= 
H_0-\sum_{i\ell}\left[\left(\frac{U}{3}+\frac{J_H}{4}\right)(\vec{S}_{i\ell}^d+\vec{S}_{i\ell}^f)^2+\left(\frac{U}{3}-\frac{J_H}{4}\right)(\vec{S}_{i\ell}^d-\vec{S}_{i\ell}^f)^2\right].
    \end{aligned}
\end{equation}
To enable mean field decoupling, we introduce two auxiliary fields $\vec{N}_{i\ell}$ and $d\vec{N}_{i\ell}$ and decouple the last line in the above equation via Hubbard-Stratonovich transformation, which leads to the following mean field Hamiltonian
\begin{equation}
    \begin{aligned}
        H'&=H_0+\sum_{i\ell}{a_+\vec{N}^2}
        +{a_-d\vec{N}^2}
-2(-1)^{i+\ell}\left[a_+\vec{N}\cdot(\vec{S}^d_{i\ell}+\vec{S}^f_{i\ell})
        +a_- d\vec{N}\cdot (\vec{S}^d_{i\ell}-\vec{S}^f_{i\ell})\right]\\
        &=H_0+\sum_{i\ell}{a_+\vec{N}^2}+{a_- d\vec{N}^2}-
        2(-1)^{i+\ell}
        \left[(a_+\vec{N}+a_-d\vec{N})\cdot \vec{S}^d_{i\ell}+(a_+\vec{N}-a_-d\vec{N})\cdot \vec{S}^f_{i\ell}
        \right]
    \end{aligned}
\end{equation}
where we define 
\begin{eqnarray}
{\vec{N}=(-1)^{i+\ell}\braket{\vec{S}^d_{i\ell}+\vec{S}^f_{i\ell}}, 
    \quad d\vec{N}=(-1)^{i+\ell}\braket{\vec{S}^d_{i\ell}-\vec{S}^f_{i\ell}}. 
    }
\end{eqnarray}
and
\begin{equation}
    a_+=\frac{U}{3}+\frac{J_H}{4}, ~ a_-=\frac{U}{3}-\frac{J_H}{4}.
\end{equation}
The physical meaning of $N$, $dN$ is the orbital summed AFM order, and their differences.

The spin operators are expressed via the fermions
\begin{equation}
    \vec{S}^d_{i\ell}=\frac{1}{2}\sum_{\alpha\beta}d^\dagger_{i\ell \alpha}\vec{\sigma}_{\alpha\beta} d_{i\ell\beta}, ~~~ \vec{S}^f_{i\ell}=\frac{1}{2}\sum_{\alpha\beta}f^\dagger_{i\ell \alpha}\vec{\sigma}_{\alpha\beta} f_{i\ell\beta}.
\end{equation}
Without  loss of generality, we assume the magnetic order is collinear and we choose 
\begin{equation}
\vec{N}=N\hat z, ~~~ d\vec{N}=dN\hat z.
\end{equation}
For a low spin AFM, we adopt the ansatz 
\begin{equation}
    a_+ N=a_- dN,
\end{equation}
which guarantees $\vec{S}_{i\ell}^f$ does not enter in the mean-field sense. It is also consistent with the $J_H\to0$ and hence $a_{+}=a_{-}$ limit where only $d_{x^2-y^2}$ orbital forms AFM and hence low spin.

For a high spin AFM, we adopt the ansatz 
\begin{equation}
    dN=0.
\end{equation}
This is consistent with the large $J_H$ limit where both spins are locked through Hund's coupling. In this limit if we keep $a_-\to 0^+$, which means still $\frac{U}{3}>\frac{J_H}{4}$, a small amount of $dN$ would be penalized by a lot of energy cost so $dN=0$ is a good choice to minimize the ground state energy.
Below we discuss these two cases separately.

\subsection{1. Low spin AFM} 
\label{sub:low_spin_afm}
\begin{figure}
\includegraphics[width=0.6\linewidth]{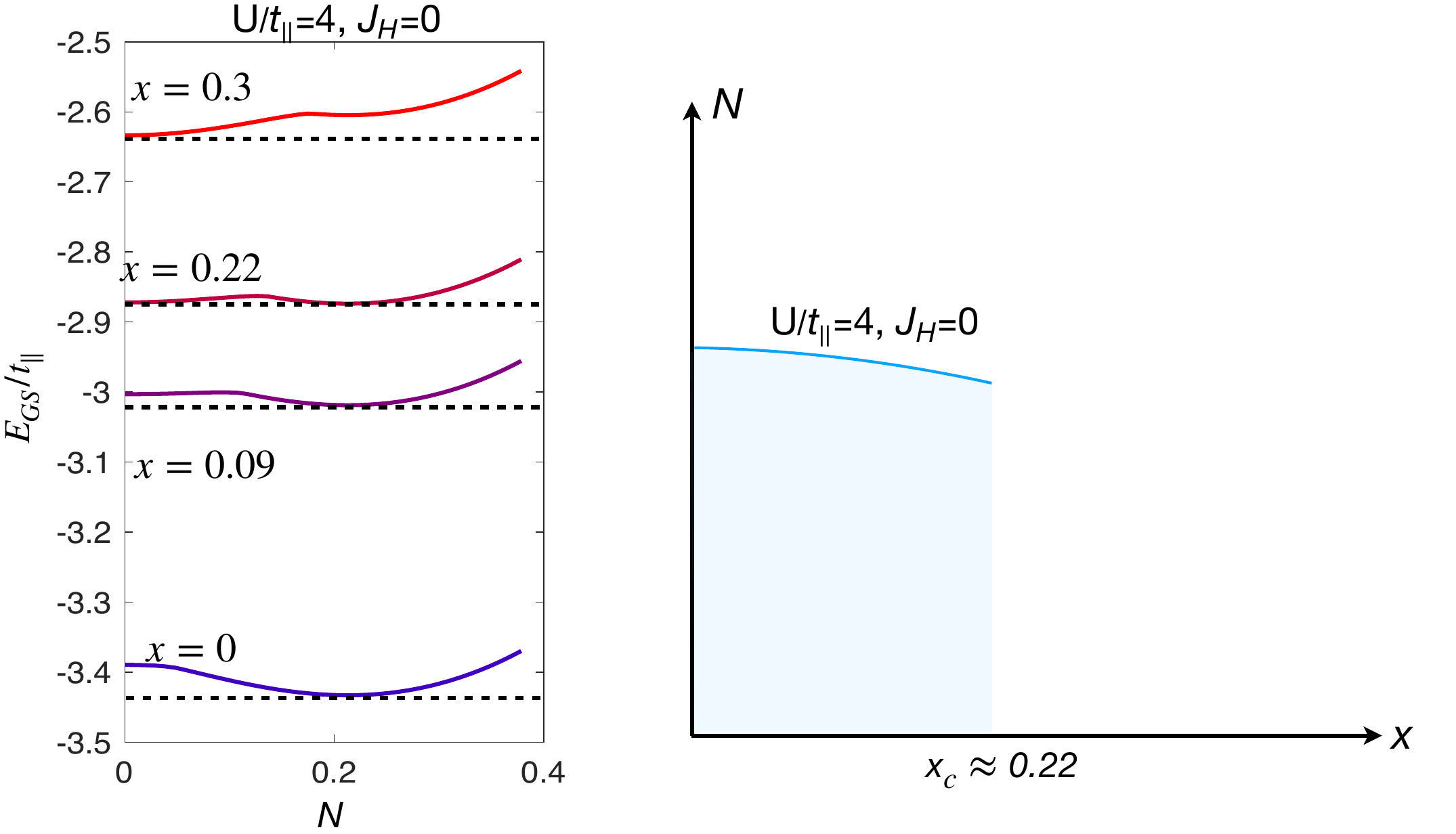}
    \caption{HF calculation result of the AFM order for a single layer single orbital Hubbard model, obtained in the limit $J_H=0$. $x$ here is the hole density such that $x=0$ corresponds to half filling ($d^8$) and $x=1$ corresponds to empty ($d^7$). }\label{fig:decouplelimit}
\end{figure}

We first discuss the case when $x^2-y^2$ and $z^2$ orbitals are completely decoupled, which is in the limit of $J_H\to0$. In this case, the low spin AFM order from $x^2-y^2$ orbitals is simply the well known AFM order in the the single layer single orbital square lattice Hubbard model. For concreteness, we choose the NN and NNN hopping the same as in Eq., and hole-dope the system starting from half filling.
In Fig.\ref{fig:decouplelimit} we show a result from HF calculation for $U=4t$ Hubbard model, with a focus on how the AFM evolves under hole concentration. Here we use $x$ to denote the average hole density, such that $x=0$ corresponds to electron half filling, and $x=1$ corresponds to no electron state.  For this particular strength of interaction, the AFM order disappears through a first order phase transition at $x=0.22$.

We now show that the same HF calculation with the two orbitals coupled are
It is convenient to introduce a new fermion basis $\Psi_{\vec{k}}=(d_{\vec{k}\uparrow},d_{\vec{k}+\vec{Q}\uparrow},d_{\vec{k}\downarrow},d_{\vec{k}+\vec{Q}\downarrow},f_{\vec{k}\uparrow},f_{\vec{k}+\vec{Q}\uparrow},f_{\vec{k}\downarrow},f_{\vec{k}+\vec{Q}\downarrow})^T$ such that the Hamiltonian can be rewritten as
\begin{equation}
    \begin{aligned}
        H'=&\frac{1}{2}\sum_{\vec{k}}\Psi^\dagger_{\vec{k}}\begin{pmatrix}
        \varepsilon_d(\vec{k}) & 2a_+N & 0 & 0 & 0 & 0 & 0 & 0\\
        2a_+N & \varepsilon_d(\vec{k}+\vec{Q}) & 0 & 0 & 0 & 0 & 0 & 0\\
        0 & 0 & \varepsilon_d(\vec{k}) & -2a_+N & 0 & 0 & 0 & 0\\
        0 & 0 & -2a_+N & \varepsilon_d(\vec{k}+\vec{Q}) & 0 & 0 & 0 & 0\\
        0 & 0 & 0 & 0 & \varepsilon_f(\vec{k}) & 0 & 0 & 0\\
        0 & 0 & 0 & 0 & 0& \varepsilon_f(\vec{k}+\vec{Q}) & 0 & 0\\
        0 & 0 & 0 & 0 & 0 & 0 & \varepsilon_f(\vec{k}) & 0\\
        0 & 0 & 0 & 0 & 0 & 0 & 0 & \varepsilon_f(\vec{k}+\vec{Q})\\
    \end{pmatrix}\Psi_{\vec{k}}+{\mathcal{N}N^2}\left(a_++\frac{a_+^2}{a_-}\right).
    \end{aligned}\label{eq:lowspin1}
\end{equation}
where $\mathcal{N}$ is the total number of the lattice sites with both layers included. Here, we use $\vec{Q}=(\pi,\pi,\pi)$. The summation of $\vec{k}$ is over the whole BZ so that we include a factor of $\frac{1}{2}$. After diagonalizing the Hamiltonian we obtain for the Hamiltonian density
\begin{equation}
    \mathcal{H}'=\frac{1}{2\times2}\sum_{n=1}^8\sum_{k_z=0,\pi}\int \frac{d^2\vec{k}}{(2\pi)^2}E_{n}(\vec{k})\psi^\dagger_{n,\vec{k}\sigma}\psi_{n,\vec{k}\sigma}+{N^2}\left(a_++\frac{a_+^2}{a_-}\right)
\end{equation}
where $E_{n}(\vec{k})$ is the $n$-th eigenvalue from diagonalizing the kernel matrix in Eq.\eqref{eq:lowspin1}. The additional factor of $\frac{1}{2}$ comes from the fact that total number of sites $\mathcal{N}$ contains two layers.
Since all these are fermions, the ground state energy density is given by
\begin{equation}
    E_{GS}(N)=\frac{1}{4}\sum_{n=1}^8\sum_{k_z=0,\pi}\int \frac{d^2\vec{k}}{(2\pi)^2}E_{n}(\vec{k})\Theta[-E_{n}(\vec{k})]+{N^2}\left(a_++\frac{a_+^2}{a_-}\right),
\end{equation}

\subsection{2. High spin AFM} 
\label{sub:high_spin_afm}
We now discuss the high spin case by setting $dN=0$, and we still use the fermion basis $\Psi_{\vec{k}}$ introduced above, such that the high spin AFM mean field Hamiltonian is written as
\begin{equation}
    \begin{aligned}
        H'=&\frac{1}{2}\sum_{\vec{k}}\Psi^\dagger_{\vec{k}}\begin{pmatrix}
        \varepsilon_d(\vec{k}) & a_+N & 0 & 0 & 0 & 0 & 0 & 0\\
        a_+N & \varepsilon_d(\vec{k}+\vec{Q}) & 0 & 0 & 0 & 0 & 0 & 0\\
        0 & 0 & \varepsilon_d(\vec{k}) & -a_+N & 0 & 0 & 0 & 0\\
        0 & 0 & -a_+N & \varepsilon_d(\vec{k}+\vec{Q}) & 0 & 0 & 0 & 0\\
        0 & 0 & 0 & 0 & \varepsilon_f(\vec{k}) & a_+N & 0 & 0\\
        0 & 0 & 0 & 0 & a_+N& \varepsilon_f(\vec{k}+\vec{Q}) & 0 & 0\\
        0 & 0 & 0 & 0 & 0 & 0 & \varepsilon_f(\vec{k}) & -a_+N\\
        0 & 0 & 0 & 0 & 0 & 0 & -a_+N & \varepsilon_f(\vec{k}+\vec{Q})\\
    \end{pmatrix}\Psi_{\vec{k}}+a_+{\mathcal{N}N^2}.
    \end{aligned}\label{eq:lowspin1}
\end{equation}
Here, we use $\vec{Q}=(\pi,\pi,\pi)$.
Similarly, the ground state energy density is given by
\begin{equation}
    \begin{aligned}
         E_{GS}(N)=\frac{1}{4}\sum_{n=1}^8\sum_{k_z=0,\pi}\int \frac{d^2\vec{k}}{(2\pi)^2}E_{n}(\vec{k})\Theta[-E_{n}(\vec{k})]+a_+{N^2}
    \end{aligned}
\end{equation}
It is easy to modify the above Hamiltonian and ground state energy by including also $dN$ into calculation. We have confirmed that the ground state energy is always minimized at $dN=0$ for large $J_H$.

\begin{figure}
    \centering
    \includegraphics[width=0.35\linewidth]{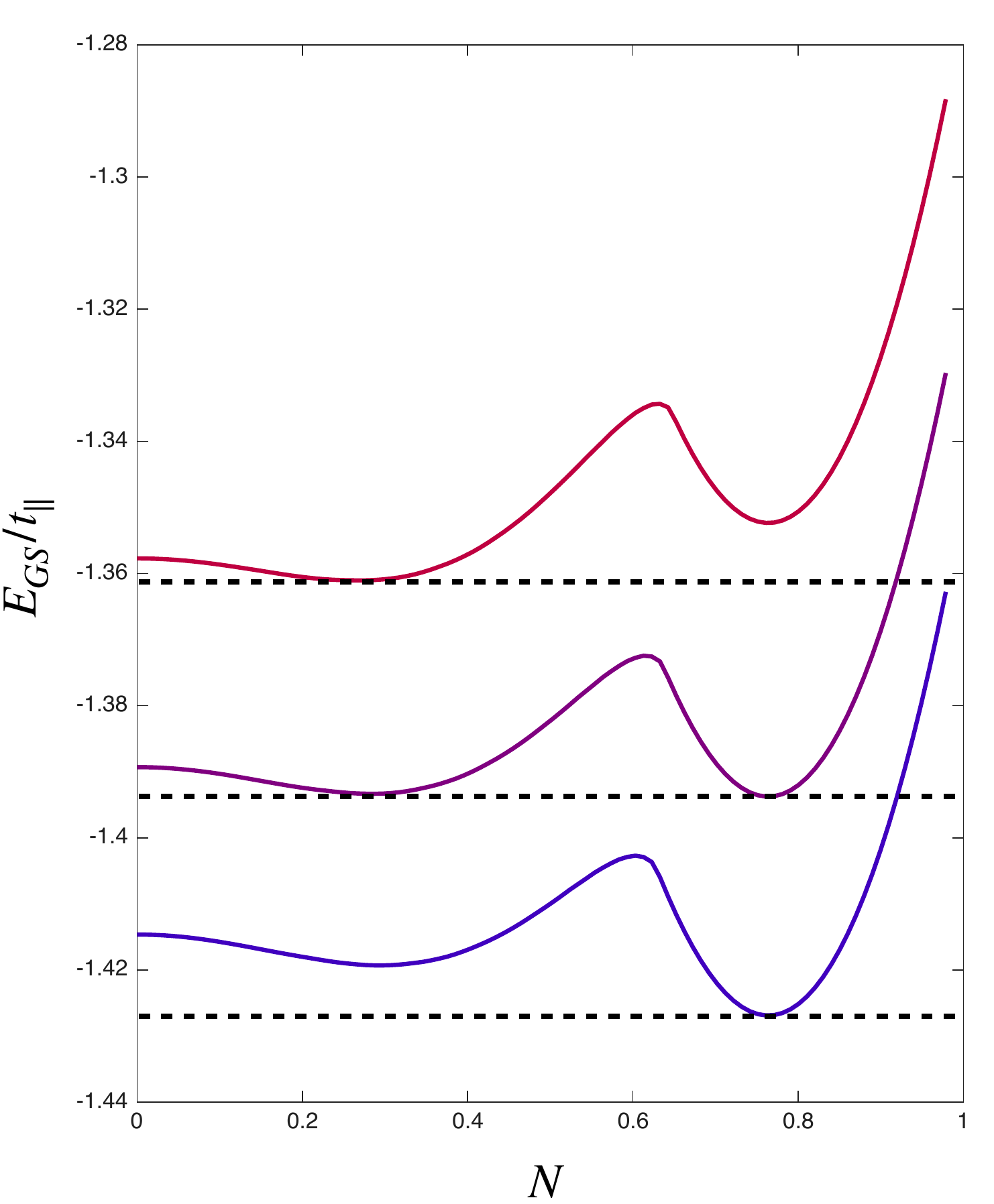} 
    \caption{{{Ground state energy as a function of $N$ for the high-spin case. Here we only present three curves near the critical doping where the discontinuous jump occurs. It is apparent that the discontinuity in $N$, i.e. the global minimum, originates from the fact that there are two local mimina in the energy profile.}} }
    \label{fig:R4}
\end{figure}

An example of the ground state energy is shown in Fig.\ref{fig:R4}. Reading the global minimum from these curves determines Fig.4(b) in the main text. Here we only show three curves near the vicinity of the discontinuity shown in Fig.4(b) as the sudden jump of $N$. From Fig.\ref{fig:R4} we clearly see that the such a sudden jump is intimately related to the presence of two local minima in the energy profile. As $x$ increase, the global minimum shifts from one of the two local minima to the other in a first order transition manner.

\end{document}